\documentclass{amsart}
\usepackage[margin=1in]{geometry}

\usepackage[utf8]{inputenc} 
\usepackage[T1]{fontenc}    
\usepackage{hyperref}       
\usepackage{url}            
\usepackage{booktabs}       
\usepackage{amsfonts}       
\usepackage{nicefrac}       
\usepackage{microtype}      
\usepackage{lipsum}		
\usepackage{graphicx}
\usepackage[square,sort,comma,numbers]{natbib}
\usepackage{doi}
\usepackage{placeins}

\usepackage{amsmath,amsfonts,amssymb,mathtools,amsthm}
\usepackage{blkarray,booktabs,multirow}
\usepackage{hyperref,cleveref}
\usepackage{graphicx,xcolor}
\usepackage[margin=0.75in,labelfont=bf,labelsep=period]{caption}
\usepackage{subfig}
\usepackage{braket}
\usepackage{arydshln}
\usepackage{myqcircuit}
\usepackage{comment}
\usepackage{thmtools}
\usetikzlibrary{calc}
\usetikzlibrary{positioning}
\usetikzlibrary{automata,arrows,matrix,quotes}

\usepackage{adjustbox}
\usepackage[edges]{forest}
\usepackage{listings}

\numberwithin{equation}{section}

\newcommand{\hlfpi}{\frac{\pi}{2}}

\usepackage{braket}
\usepackage{amsthm}

  \usepackage{tikz}
  \usepackage{pgfplots}
  \usetikzlibrary{calc}
  \usetikzlibrary{positioning}
  \usetikzlibrary{plotmarks}
  \usetikzlibrary{automata,arrows,matrix,quotes}
  \pgfplotsset{
    compat=newest,
    table/header=false,
    tick label style={font=\scriptsize},
    label style={font=\scriptsize},
    legend style={font=\scriptsize},
    legend cell align=left,
    colormap={parula}{
      rgb255=(53,42,135)
      rgb255=(15,92,221)
      rgb255=(18,125,216)
      rgb255=(7,156,207)
      rgb255=(21,177,180)
      rgb255=(89,189,140)
      rgb255=(165,190,107)
      rgb255=(225,185,82)
      rgb255=(252,206,46)
      rgb255=(249,251,14)
    }
  }
  \pgfplotsset{
    myColOne/.style={myblue},
    myColTwo/.style={myred},
    myColThr/.style={myorange},
    myColFou/.style={mypurple},
    myColFiv/.style={mygreen},
    myColSix/.style={mylightblue},
    myColSev/.style={mydarkred}
  }

\newcommand{\eps}{\epsilon}

\newcommand{\abs}[1]{\left|#1\right|}

\newcommand{\norm}[1]{\lVert#1\rVert}

\newcommand{\bbm}{\begin{bmatrix}}
	\newcommand{\ebm}{\end{bmatrix}}
\newcommand{\RR}{\mathbb{R}}
\newcommand{\CC}{\mathbb{C}}

\newcommand{\ZZ}{\mathbb{Z}}

\newcommand{\Z}{\Xi}

\newcommand{\diag}{\mathrm{diag}}

\newcommand{\half}{\frac{1}{2}}

\newcommand{\CNOT}{\mathrm{CNOT}}
\newcommand{\SWAP}{\mathrm{SWAP}}

\newcommand{\pjp}{\psi_{2Bj+p}}
\newcommand{\fpjp}{\widehat{\pjp}}
\newcommand{\cpjp}{a_{2Bj+p}}
\newcommand{\inner}[1]{\langle #1 \rangle}
\newcommand{\snb}[1]{S_{\mathrm{G}(#1)}}

\newcommand{\halfpi}{\frac{\pi}{2}}
\newcommand{\halfi}{\frac{i}{2}}
\newcommand{\tg}{T_{\mathrm{G}}}
\newcommand{\vg}{V_{\mathrm{G}}}
\newcommand{\qpm}[1]{Q_{\mathrm{G}(#1)}}

\newcommand{\rpm}[1]{G_{\mathrm{G}(#1)}}
\newcommand{\singleodd}{\hat{K}_o}
\newcommand{\singleeven}{\hat{K}_e}
\newcommand{\twoodd}{K_o}
\newcommand{\twoeven}{K_e}
\newcommand{\ugs}{U_{\mathrm{GS}}}
\newcommand{\ugb}{U_{\mathrm{GB}}}
\newcommand{\cwjp}{a_{j,p}}
\newcommand{\wjp}{\psi_{j,p}}
\newcommand{\fwjp}{\widehat{\psi_{j,p}}}
\newcommand{\shf}[1]{S_{\mathrm{W}(#1)}}
\newcommand{\rz}{R_z}
\newcommand{\uws}{U_{\mathrm{WS}}}
\newcommand{\trans}[1]{G_{\mathrm{W}(#1)}}
\newcommand{\swaponethree}{P_{1,3}}
\newcommand{\cwjpone}{a_{1,p}}

\newcommand{\ketbra}[2]{\ket{#1}\!\bra{#2}}
\newcommand{\qpi}{\frac{\pi}{4}}
\newcommand{\hpi}{\frac{\pi}{2}}
\newcommand{\wtg}{T_{\mathrm{W}}}
\newcommand{\uwb}{U_{\mathrm{WB}}}
\newcommand{\cscale}{a_{n+1,0}}
\newcommand{\wq}[1]{Q_{\mathrm{W}(#1)}}
\newcommand{\wk}[1]{K_{#1}}
\newcommand{\wlj}[1]{L_{#1}}
\newcommand{\wrj}[1]{R_{#1}}
\newcommand{\jone}{\underbrace{1\cdots1}_{(j-1)\text{ of 1's}}\!\!\!\!\!1}
\newcommand{\jzero}{\underbrace{0\cdots0}_{(j-1)\text{ of 0's}}\!\!\!\!\!1}

\newcommand{\rd}{\mathrm{d}}
\usepackage{upgreek}

\title{Quantum Wave Packet Transforms with compact frequency support}
\author[]{Hongkang Ni} \address[Hongkang Ni]{Stanford University,
  Stanford, CA 94305} \email{hongkang@stanford.edu}

\author[]{Lexing Ying} \address[Lexing Ying]{Stanford University,
  Stanford, CA 94305} \email{lexing@stanford.edu}

\thanks{The work of L.Y. is partially supported by the National Science Foundation under awards DMS-2011699 and DMS-2208163.}


\begin{document}

\begin{abstract}
  Different kinds of wave packet transforms are widely used for extracting multi-scale structures in signal processing tasks. This paper introduces the quantum circuit implementation of a broad class of wave packets, including Gabor atoms and wavelets, with compact frequency support. Our approach operates in the frequency space, involving reallocation and reshuffling of signals tailored for manipulation on quantum computers. The resulting implementation is different from the existing quantum algorithms for spatially compactly supported wavelets and can be readily extended to quantum transforms of other wave packets with compact frequency support.
\end{abstract}

\maketitle

\section{Introduction}
Various types of wave packets, such as wavelets \cite{mallat1999wavelet}, Gabor atoms, curvelets \cite{candes2005curvelet}, wavelet packets \cite{lintner2004solving}, and wave atoms \cite{demanet2007wave}, have been extensively explored and applied to many applications. In contrast to the Fourier and the Dirac bases, these wave packets exhibit locality in both the spatial and frequency domains, along with a controlled level of smoothness, making them advantageous for image or other signal processing tasks. Additionally, research has demonstrated that the structure of these wave packet bases offers sparse representations for some differential operators, enabling their use as effective preconditioners for partial differential equations and facilitating the development of rapid algorithms \cite{candes2003curvelets, demanet2009wave}.

There has been emerging interest in quantum algorithms utilizing wavelet or other wave packet transformations. Research works have demonstrated the applicability of quantum wavelet transforms in various areas of image processing \cite{Yan2016a, Zhao2022duallevel}, including image compression and encryption \cite{Zhou2020novel}, image denoising \cite{Chakraborty2020an}, and image watermarking \cite{Yu2023adaptive, Song2013a, Mu2023an, Heidari2017a}. Furthermore, quantum wavelet transform has found applications in diverse fields such as quantum steganography \cite{Gao2022a}, quantum-to-classical data decoding \cite{Jeng2023improving}, and signal reconstruction \cite{Moshtaghpour2020close}. Notably, quantum wavelet and curvelet transforms have also been found valuable in encoding differential operators \cite{Kiani2022quantum, Bagherimehrab2023fast} and in the domain of quantum sampling \cite{Liu2009quantum}.

Similar to the Fourier transform, quantum computing presents the potential for exponential speedup for wave packet transforms. The implementations of various quantum wavelet transforms have been explored. Earlier studies included well-known wavelets such as the Haar wavelet and Daubechies D4 wavelet \cite{fijany1999quantum, Li2019quantum_multi}. Research has been extended to higher-order wavelets \cite{Bagherimehrab2023efficient} 
and a more general setting \cite{Zhang2022wavelet}. The exploration of multi-dimensional quantum wavelet transform has also been undertaken \cite{Li2023threedimensional, Li2022multilevel, Li2018the_multi}. Wavelet packet transforms have been addressed in studies like \cite{klappenecker1999wavelets, ChaurraGutierrez2023qist}. However, there is a noticeable gap in research concerning other types of wave packet transforms. \cite{Liu2009quantum} touched upon the multi-dimensional curvelet transform and its potential applications, but the discussion remains somewhat conceptual without a detailed exploration of its implementation.


\subsection{Main contributions}


Previous studies on quantum wavelet transforms have mostly utilized finite-size filters in the spatial domain, which are only applicable to spatially compactly supported wavelets. On the other hand, wave packet constructions with compact frequency support, such as the Meyer wavelet, exhibit favorable characteristics in the frequency domain. As far as we know, no existing algorithm has explored their implementations on quantum computers. Our contribution lies in operating within the frequency domain and executing the implementation of such wave packet constructions.

We first consider the Gabor atoms of uniform frequency partitioning, with both sharp and blended frequency windows. The implementation given here is the first one on quantum computers and also serves as a preparation for more advanced wave packets. 

Next, we consider the wavelets of multiscale frequency partitioning with both sharp and blended frequency windows. The two examples studied in detail are Shannon wavelets and Meyer wavelets. We show that these wavelets may be realized with no more than three ancilla qubits, thanks to their favorable forms in the frequency domain. In contrast, for the wavelets with compact spatial support, the number of ancilla qubits typically grows with the order of the wavelet \cite{Bagherimehrab2023efficient}. 

\subsection{Background on Gabor atoms and wavelets}\label{sec: background}

Different tilings of phase-space diagrams result in different basis sets, which can be collectively called wave packets. This work focuses on two such wave packets: Gabor atoms and wavelets.

\begin{figure}[!htb]
    \centering
    \subfloat[\label{fig: wavepacket_tile}]{
    \begin{minipage}[c]{0.45\textwidth}
        \centering
        \includegraphics[width=0.95\textwidth]{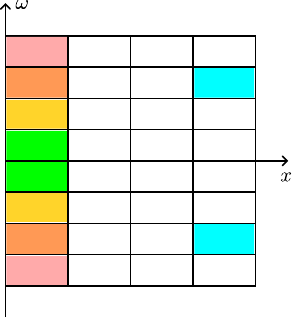}
    \end{minipage}
    }
    \subfloat[\label{fig: wavelet_tile}]{
    \begin{minipage}[c]{0.45\textwidth}
        \centering
        \includegraphics[width=0.95\textwidth]{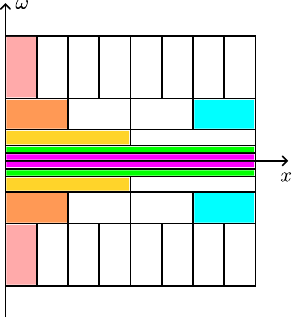}
    \end{minipage}
    }
    \caption{(a) Illustration for Gabor atom tiling. Each pair of tiles symmetric with respect to the $x$ axis represents the essential support of a basis function $\uppsi_{j,p}$. The red, orange, yellow, and green tiles represent $\uppsi_{4,0}$, $\uppsi_{3,0}$, $\uppsi_{2,0}$, and $\uppsi_{1,0}$ respectively. All tiles with the same $\omega$ are at the same level $j$. For instance, the blue tile represents $\uppsi_{3,3}$. (b) Illustration for wavelet tiling. Each pair of tiles symmetric with respect to the $x$ axis represents the essential support of a basis function $\uppsi_{j,p}$. This diagram shows the truncation up to $n=4$. The red, orange, yellow, green, and blue tiles represent $\uppsi_{1,0}$, $\uppsi_{2,0}$, $\uppsi_{3,0}$,  $\uppsi_{4,0}$, and $\uppsi_{2,3}$ respectively. The purple tiles represent the scaling function.}
    \label{fig: tile}
\end{figure}

\subsubsection{Gabor atoms}
Let us first consider a uniform phase-space tiling, as illustrated in \Cref{fig: wavepacket_tile}. The basis functions given by this uniform tiling are called Gabor atoms. A basis function can be indexed by its frequency level $j\in\ZZ_{>0}$ and its space position $p\in\ZZ$. 
As the simplest example, the Gabor atoms with {\em sharp} frequency windows are given by
\begin{equation}\label{eq: Shannon wavepacket cont}
    \widehat{\uppsi_{j,p}}(\omega) = \sqrt{\frac{A}{2\pi}}e^{iA\omega p}\chi_{[j\frac{\pi}{A},(j+1)\frac{\pi}{A}]}(|\omega|),
\end{equation}
where $\chi$ stands for the indicator function. Here, we adopt the convention
\[
\hat{\uppsi}(\omega) = \frac{1}{\sqrt{2\pi}}\int_{\RR}\uppsi(x)e^{i\omega x}\rd x,
\]
for continuous Fourier transform, and $A\in\ZZ_{>0}$ is a scaling parameter that can be arbitrarily chosen. Though with perfect frequency localization, these atoms have slow spatial decay. 

A smooth frequency windowing is needed for Gabor atoms with better spatial localization. For this purpose, we can define a bump function
\begin{equation}\label{eq: gx}
    g(s) = 
\begin{cases}
 \cos(\frac{\pi}{2} \beta(s/\pi))& \text{if $-\pi< s <\pi$,} \\
0 & \text{otherwise,}
\end{cases}
\end{equation}
where the function $\beta(s)$ should satisfy 
\begin{equation}\label{eq: constraints on beta}
    \beta(s)+\beta(1-s)=1 \text{ and }\beta(-s) = \beta(s).
\end{equation}
The bump function $g(s)$ can be viewed as a smoothed version of the characteristic function $\chi_{[-\halfpi,\halfpi]}$, and the order of smoothness is determined by $\beta$. 
The Gabor atoms with {\em blended} frequency windows are given by 
\begin{equation}\label{eq: Meyer wavepacket cont}
    \widehat{\uppsi_{j,p}}(\omega) = \sqrt{\frac{A}{2\pi}}e^{iA\omega p}\left(e^{-i\half(A\omega-(j+\half)\pi)} g\left(A\omega-(j+\half)\pi\right) + e^{-i\half(A\omega+(j+\half)\pi)} g\left(A\omega+(j+\half)\pi\right)\right).
\end{equation}


\subsubsection{Wavelets}
For wavelets, the phase-space tiling is given as in \Cref{fig: wavelet_tile}. More specifically, a wavelet basis is given by a function $\uppsi:\RR\to \CC$ and its multiscale counterparts:
\[
\uppsi_{j,p}(x) = \frac{1}{\sqrt{2^j}}\uppsi\left(\frac{x-2^jp}{2^j}\right),
\]
such that $\{\uppsi_{j,p}\}_{(j,p)\in\ZZ^2}$ form an orthonormal basis of $L^2(\RR)$. As a result, their Fourier transforms
\begin{equation}\label{eq: wavelet cont}
\widehat{\uppsi_{j,p}}(\omega) = \sqrt{\frac{2^j}{2\pi}}e^{i2^j\omega p}\hat{\uppsi}(2^j\omega)
\end{equation}
also form an orthonormal basis of $L^2(\RR)$.  When the frequency window $\hat{\uppsi}(\omega)$ is a {\em sharp} indicator function
\begin{equation}\label{eq: Shannon wavelet cont}
\hat{\uppsi}(\omega) = \chi_{[-2\pi,-\pi]\cup[\pi,2\pi]}(\omega) = \chi_{[\pi,2\pi]}(|\omega|),
\end{equation}
the resulting basis functions are the Shannon wavelets. 

We may also consider smooth {\em blended} frequency windows, for example,
\begin{equation}\label{eq: defi phi omega}
    \hat{\uppsi}(\omega) = \begin{cases}
        e^{-i\frac{\omega}{2}}g\left(\frac{3\omega}{2}-2\pi\right)& \text{if $\frac{2\pi}{3}\le \omega\le\frac{4\pi}{3}$,}\\
        e^{-i\frac{\omega}{2}}g\left(\frac{3\omega}{4}-\pi\right)& \text{if $\frac{4\pi}{3}\le \omega\le\frac{8\pi}{3}$,}\\
        0& \text{if $0\le \omega\le\frac{2\pi}{3}$ or $\omega\ge\frac{8\pi}{3}$,}\\
        \hat{\uppsi}(-\omega)^*& \text{if $\omega<0$,}\\
    \end{cases}
\end{equation}
where $\hat{\uppsi}(-\omega)^*$ means the conjugate of complex number $\hat{\uppsi}(-\omega)$. This results in the famous Meyer wavelets \cite[Chapter 7.2.2]{mallat1999wavelet}.





\subsection{Notations and conventions}\label{sec: notation}
Let us first summarize the notations and conventions used in this paper, which are compatible with the standard notations in quantum computing literature. We use $\otimes$ to denote the tensor product and $\oplus$ for the direct product of matrices. Occasionally, $\oplus$ is also used as the notation of modulo 2 plus in bit-wise operations, but the meaning should be clear based on the context. The notation $[M]$ stands for the set $\{0,1,\ldots,M-1\}$. The matrix $I_M$ means the identity matrix of dimension $M$. We use $z^*$ to represent the conjugate of a complex number $z$, and $A^\dagger$ the conjugate transpose of matrix $A$.

We adopt Dirac's notation $\ket{\cdot}$ to represent a quantum state as a column vector and $\bra{\cdot}$ as its conjugate transpose. A qubit lives in the space $\CC^2=\mathrm{span}\{\ket{0},\ket{1}\}$, where $\ket{0}$ and $\ket{1}$ represent $[1,0]^T$ and $[0,1]^T$, respectively. For a non-negative integer $x$ with binary representation $x = (x_{m-1}x_{m-2}\cdots x_0)_2$, we denote the $m$-qubit state
\begin{equation}\label{eq: binary repr}
    \ket{x} := \ket{x_{m-1}x_{m-2}\cdots x_0} = \ket{x_{m-1}}\ket{x_{m-2}}\cdots \ket{x_0} = \ket{x_{m-1}}\otimes\ket{x_{m-2}}\otimes\cdots \otimes\ket{x_0}.
\end{equation}
In this convention, we can identify any unit vector $f\in\CC^M$ as an $m$-qubit state $\sum_{x=0}^{M-1} f(x)\ket{x}$. Due to the basic assumptions of quantum mechanics, we always assume the vectors have norm 1, and all operators we consider are unitary matrices correspondingly.

The unitary manipulations are often drawn as boxes in the quantum circuits, where we use horizontal wires to represent qubits. The ordering convention we are using here is the higher digit of the binary representation \eqref{eq: binary repr} will correspond to the upper qubit when drawing the quantum circuit. More examples of quantum circuits can be found in quantum computing textbooks \cite{lin2022lecture, NielsenChuang2000}. In the present paper, we denote $n=\log_2 N$ as the number of qubits utilized for implementing the wave packet transform in the space $\mathbb{C}^N$. Additionally, $M=2^m$ is often used for the dimension of matrices acting as intermediate steps, with the value of $M$ left temporarily unspecified.

\begin{figure}[hbtp]
\centering
\subfloat[\label{fig: cnot1}]{%
\begin{minipage}[c]{0.24\textwidth}
\centering
\begin{tikzpicture}[on grid]
\pgfkeys{/myqcircuit, layer width=7.5mm, row sep=5mm, source node=qwsource}
\newcommand{\qwstart}{1}
\newcommand{\qwend}{3}
\qwire[start node=qw1s, end node=qw1, style=thin, index=1, start layer=\qwstart, end layer=\qwend]
\qwire[start node=qw2s, end node=qw2, style=thin, index=2, start layer=\qwstart, end layer=\qwend]
\cnot[layer=1, control index=1, target index=2]
\renewcommand{\qwstart}{4}
\renewcommand{\qwend}{6}
\qwire[start node=qw1s, end node=qw1e, style=thin, index=1, start layer=\qwstart, end layer=\qwend]
\qwire[start node=qw2s, end node=qw2e, style=thin, index=2, start layer=\qwstart, end layer=\qwend]
\singlequbit[style=gate, layer=4, index=2, node=H5, label=$X$]
\control[layer=4, index=1, target node=H5, node = ctrl1, style=controlon]
\node at ($0.5*(qw1)+0.5*(qw2)+(0.4,0)$) {$=$};
\end{tikzpicture}%
\end{minipage}%
}
\subfloat[\label{fig: cnot0}]{%
\begin{minipage}[c]{0.24\textwidth}
\centering
\begin{tikzpicture}[on grid]
\pgfkeys{/myqcircuit, layer width=7.5mm, row sep=5mm, source node=qwsource}
\newcommand{\qwstart}{1}
\newcommand{\qwend}{3}
\qwire[start node=qw1s, end node=qw1e, index=1, start layer=\qwstart, end layer=\qwend]
\qwire[start node=qw2s, end node=qw2e, index=2, start layer=\qwstart, end layer=\qwend]
\cnot[layer=1, control index=1, target index=2, style=controloff]
\end{tikzpicture}%
\end{minipage}%
}
\subfloat[\label{fig: swap}]{%
\begin{minipage}[c]{0.24\textwidth}
\centering
\begin{tikzpicture}[on grid]
\pgfkeys{/myqcircuit, layer width=7.5mm, row sep=5mm, source node=qwsource}
\newcommand{\qwstart}{1}
\newcommand{\qwend}{3}
\qwire[start node=qw1s, end node=qw1e, index=1, start layer=\qwstart, end layer=\qwend]
\qwire[start node=qw2s, end node=qw2e, index=2, start layer=\qwstart, end layer=\qwend]
\swapgate[layer=1, first index=1, second index=2]
\end{tikzpicture}%
\end{minipage}%
}
\subfloat[\label{fig: multicnot1}]{%
\begin{minipage}[c]{0.24\textwidth}
\centering
\begin{tikzpicture}[on grid]
\pgfkeys{/myqcircuit, layer width=7.5mm, row sep=5mm, source node=qwsource}
\newcommand{\qwstart}{1}
\newcommand{\qwend}{3}
\qwire[start node=qw1s, end node=qw1e, index=1, start layer=\qwstart, end layer=\qwend]
\qwire[start node=qw2s, end node=qw2e, index=2, start layer=\qwstart, end layer=\qwend]
\qwire[start node=qw3s, end node=qw3e, index=3, start layer=\qwstart, end layer=\qwend]
\singlequbit[style=gate, layer=1, index=3, node=H5, label=$U$]
\control[layer=1, index=2, target node=H5, node = ctrl1, style=controlon]
\control[layer=1, index=1, target node=ctrl1, style=controlon]
\end{tikzpicture}%
\end{minipage}%
}
\caption{(a) CNOT gate. (b) $\ket{0}$-controlled-NOT gate. (c) SWAP gate. (d) An example of a multi-qubit control gate $\ketbra{11}{11}\otimes U+(I_4-\ketbra{11}{11})\otimes I$. The two filled nodes indicate that $U$ is applied to the third qubit if and only if both the first two qubits are at state $\ket{1}$.}
\label{fig: cnotgate}
\end{figure}
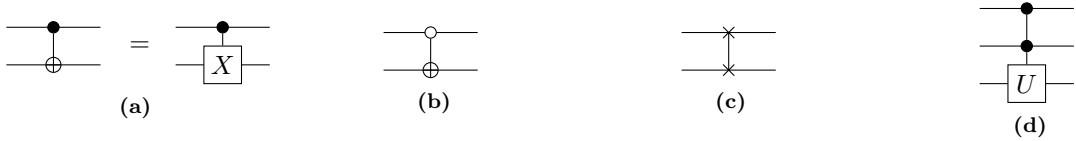

$X$, $Y$, $Z$, and $H$ represent the Pauli-$X$, Pauli-$Y$, Pauli-$Z$, and Hadamard matrices respectively, defined as
\begin{align*}
X &= \begin{bmatrix}
0 & 1 \\
1 & 0
\end{bmatrix}, &
Y &= \begin{bmatrix}
0 & -i \\
i &  \phantom{-}0     
\end{bmatrix}, &
Z &= \begin{bmatrix}
1 &  \phantom{-}0 \\
0 & -1 
\end{bmatrix}, &
H &= \frac{1}{\sqrt{2}} \begin{bmatrix}
1 & \phantom{-}1 \\
1 & -1
\end{bmatrix}.
\label{eq: basicgates}
\end{align*}
We also extensively use some basic two-qubit gates. The $\ket{1}$-controlled-NOT gate is $\ketbra{0}{0}\otimes I_2+\ketbra{1}{1}\otimes X$, and is usually just called the CNOT gate. In a quantum circuit, this gate is drawn as in \Cref{fig: cnot1}, where the filled node means only applying the lower gate when the upper qubit is $\ket{1}$. We usually use ``$\oplus$'' to represent ``$X$'' in quantum circuits. The related $\ket{0}$-controlled-NOT gate, $\ketbra{0}{0}\otimes X+\ketbra{1}{1}\otimes I_2$, is given in \Cref{fig: cnot0}, where the unfilled node has just the opposite meaning as the filled one. The swap gate $\SWAP = \ketbra{00}{00}+\ketbra{01}{10}+\ketbra{10}{01}+\ketbra{11}{11}$, which swaps the value of two qubits, is shown as \Cref{fig: swap}.  We sometimes also make use of the multi-qubit controlled gate, which can be built by single qubit gates and CNOT gates \cite{BarencoBennettCleveEtAl1995}. For instance, $\ketbra{11}{11}\otimes U+(I_4-\ketbra{11}{11})\otimes I$ is shown in \Cref{fig: multicnot1}.

An $(m+n)$-qubit unitary operator $U$ is called a $(\gamma, m, \epsilon)$-block-encoding of an $N\times N$ matrix $A$, if
$$
\left\|A-\gamma\left(\left\langle 0^m\right| \otimes I_N\right) U\left(\left|0^m\right\rangle \otimes I_N\right)\right\| \leq \epsilon.
$$
In the matrix form, a $(\gamma, m, \epsilon)$-block-encoding is a $2^{m+n}$ dimensional unitary matrix
$$
U=\left(\begin{array}{cc}
\widetilde{A} / \gamma & * \\
* & *
\end{array}\right)
$$
where $*$ can be any block matrices of the correct size, and $\|\widetilde{A}-A\| \leq \epsilon$.

The inner product of two complex vectors $f_1$ and $f_2$ is denoted by
\begin{equation}
    \braket{f_1|f_2} := \langle f_1, f_2\rangle = \sum_{t=0}^{N-1} f_1(t)f_2(t)^*.
\end{equation}
We denote $f(m_1:m_2)$ as the restriction of vector $f$ from the $m_1$-th element to the $m_2$-th element.

For any $j$ in the computational basis, the (discrete) forward Fourier transform is defined as
\[
U_{\mathrm{FT}(N)}|j\rangle=\frac{1}{\sqrt{N}} \sum_{k \in[N]} e^{\mathrm{i} 2 \pi \frac{k j}{N}}|k\rangle,
\]
which is widely used in the quantum computing literature but has a sign difference compared to the classical signal processing convention.

The indices of vectors in this paper are always considered in the modulo $N$ sense, i.e., $f(-k) = f(N-k)$. In particular, we adopt the convention $\ket{-k} = \ket{N-k}$ in the bracket notation.

 For all matrices, the subscripts ``G'' and ``W'' indicate that they are related to Gabor atoms and wavelets, respectively. The subscripts ``S'' and ``B'' stand for sharp and blended frequency windows, respectively. In particular, the four main transformations in the paper are denoted as $\ugs$, $\ugb$, $\uws$, and $\uwb$, i,e., sharp Gabor atoms, blended Gabor atoms, Shannon (sharp) wavelets, and Meyer (blended) wavelets. For the matrices that we want to specify the dimension, we write the dimension as a subscript. The italicized subscripts representing dimensions, such as $N$, $M$, and $B$, can be readily distinguished from the Roman subscripts denoting names, such as the aforementioned ``G'', ``W'', ``S'', and ``B''.

\section{Gabor atoms}\label{sec: wave packet}
This section delves into the implementation of the quantum version of the Gabor atom transform. Specifically, we will discuss the discretization and present the quantum circuits for both the sharp Gabor atoms \eqref{eq: Shannon wavepacket cont} and the blended Gabor atoms \eqref{eq: Meyer wavepacket cont}. Implementing the sharp Gabor atoms can be viewed as first reshuffling the Fourier transform $\hat{f}$ of the input and then performing an inverse Fourier transform for each sharp frequency window.  The blended Gabor atoms are more intricate due to the overlapping of neighboring packets. Our approach entails first recombining parts of $\hat{f}$ to the proper locations, which is a unitary process, and then applying the circuit of the sharp Gabor atoms. The recombining process involves the implementation of certain diagonal matrices, which will also be discussed in detail.

\subsection{Sharp Gabor atoms}\label{sec: sharp Gabor atoms}
Given the basis functions \eqref{eq: Shannon wavepacket cont}, let us introduce the discrete sharp Gabor atoms in the frequency domain. We restrict $\omega\in [-\pi,\pi)$ to grid points $\{\frac{2\pi}{N} k\}$, where $k=-N/2, \ldots, N/2-1$. Denote $B=\frac{N}{2A}$ and $b=\log_2B$. 
The discrete sharp Gabor atoms are defined in the (discrete) frequency space as 
\[
\widehat{\psi_{j,p}}(k):= \frac{1}{\sqrt{2B}}e^{2\pi i \frac{pk}{2B}}\left[ \chi_{[jB,(j+1)B)}(k) + \chi_{[-(j+1)B,-jB)}(k)\right],
\]
where $\psi_{j,p}$ is for the discrete basis function in order to distinguish from its continuous counterpart $\uppsi_{j,p}$. We also point out that the discrete and continuous basis corresponds in the sense that $\widehat{\psi_{j,p}}(k):= \sqrt{\frac{2\pi}{N}}\widehat{\uppsi_{j,p}}(\frac{2\pi}{N} k)$. Only the basis function with indices $j\in [A]$ and $p\in [2B]$ are used in the discrete setting, which is in total $N$ basis functions. Typically, we may choose $A=2^{\lfloor n/2\rfloor}$ and $B=2^{\lfloor (n-1)/2\rfloor}$ to balance the resolution of space and frequency, though other choices of $A$ and $B$ are also acceptable. To simplify indexing, we further denote $\psi_{j,p}$ as $\psi_{2Bj+p}$, thereby aligning the basis functions with indices ranging from $0$ to $N-1$. More explicitly, the Fourier coefficients of the sharp Gabor atoms basis are 
\begin{equation}\label{eq: Shannon basis}
    \fpjp(k) := \widehat{\psi_{j,p}}(k),
\end{equation}
for $j \in [\frac{N}{2B}]$, $p \in [2B]$, and $k\in\{-\frac{N}{2}, -\frac{N}{2}+1, \ldots, \frac{N}{2}-1\}$. Here, $\chi_I$ means the characteristic function on interval $I$. In some other literature, there may be an additional phase on the characteristic functions in the sharp Gabor atoms, which is not adopted here and can be viewed as a special version of the blended Gabor atoms discussed in \Cref{sec: blended Gabor atoms}. 

For a signal $f\in\CC^N$, one can expand $f$ under this basis
\begin{equation}
    f = \sum_{j\in [\frac{N}{2B}]}\sum_{p \in [2B]}\cpjp\pjp = \sum_{n=0}^{N-1}a_n\psi_n,
\end{equation}
where $\cpjp$'s are the Gabor atom coefficients. Notice that the $\pjp$ here is the \emph{discrete} Fourier inverse of \eqref{eq: Shannon basis}, which is not the direct restriction of the continuous $\uppsi_{j,p}$ in the space domain. The sharp Gabor atom transform is defined as
\begin{equation}
    \ugs:\CC^N\to \CC^N: f\mapsto a = (a_{0},a_{1},\ldots,a_{N-1})^T.
\end{equation}

\begin{figure}[!ht]
	\centering
	\includegraphics[
	width =0.7\textwidth
	]{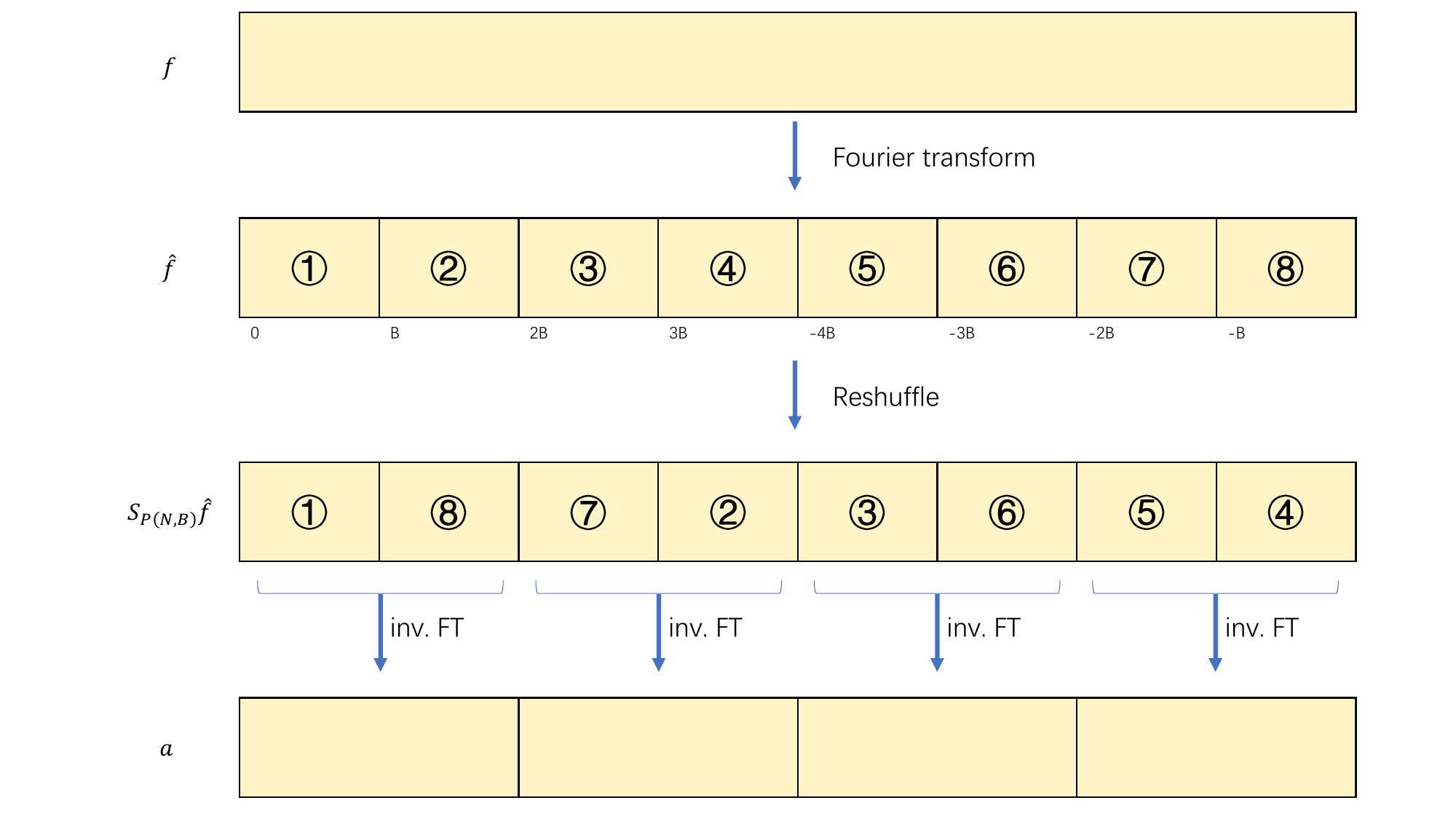}
	\caption{The process of sharp Gabor atoms transform. Reshuffle the indices in the frequency domain, and then perform an inverse Fourier transform on each of the $2B$-size blocks.}
	\label{fig: sharp Gabor atoms reshuffle}
\end{figure}

One can check that the basis given by \eqref{eq: Shannon basis} is orthonormal. Therefore, the coefficients of signal $f$ are given by the inner product
\begin{equation}\label{eq: sharp Gabor transform calculation}
    \cpjp = \inner{f,\pjp} = \inner{\hat{f},\fpjp} = \left(\sum_{k=jB}^{(j+1)B-1}+\sum_{k=-(j+1)B}^{-jB-1}\right)\frac{1}{\sqrt{2B}}e^{-2\pi i \frac{pk}{2B}}\hat{f}(k),
\end{equation}
where the second step is the Plancherel's identity. When $j$ is even, this gives
\begin{equation}
\begin{aligned}
    \cpjp &= \frac{1}{\sqrt{2B}}\left(\sum_{k=0}^{B-1}e^{-2\pi i \frac{p(k+jB)}{2B}}\hat{f}(k+jB) + \sum_{k=B}^{2B-1}e^{-2\pi i \frac{p(k-(j+2)B)}{2B}}\hat{f}(k-(j+2)B)\right) \\
    &= \frac{1}{\sqrt{2B}}\left(\sum_{k=0}^{B-1}e^{-2\pi i \frac{pk}{2B}}\hat{f}(k+jB) + \sum_{k=B}^{2B-1}e^{-2\pi i \frac{pk}{2B}}\hat{f}(k-(j+2)B)\right).
\end{aligned}    
\end{equation}
Similarly, when $j$ is odd,
\begin{equation}
    \cpjp = \frac{1}{\sqrt{2B}}\left(\sum_{k=0}^{B-1}e^{-2\pi i \frac{pk}{2B}}\hat{f}(k-(j+1)B) + \sum_{k=B}^{2B-1}e^{-2\pi i \frac{pk}{2B}}\hat{f}(k+(j-1)B)\right).   
\end{equation}
This form implies that if we permute the elements of $\hat{f}$, then $\cpjp$ can be obtained from an inverse Fourier transform of size $2B$. Specifically, let $\snb{N,B}$ be the permutation matrix such that for $j \in [\frac{N}{2B}]$, $k \in [B]$, we have
\begin{equation}\label{eq: snb cases}
\begin{aligned}
\snb{N,B}\ket{Bj+k}  &= \begin{cases}
    \ket{2Bj+k}        & \text{if $j$ is even,} \\
    \ket{2Bj+B+k}  & \text{if $j$ is odd,} 
\end{cases}\\
\snb{N,B}\ket{B(\frac{N}{B}-1-j)+k} &= \begin{cases}
    \ket{2Bj+k}     & \text{if $j$ is odd,} \\
    \ket{2Bj+B+k}   & \text{if $j$ is even.} 
\end{cases}.
\end{aligned}
\end{equation}
After acting $\snb{N,B}$ on $\hat{f}$, we have
\begin{equation}
    \cpjp = \frac{1}{\sqrt{2B}}\left(\sum_{k=0}^{2B-1}e^{-2\pi i \frac{pk}{2B}}(\snb{N,B}\hat{f})(k+2Bj)\right),   
\end{equation}
which is just an inverse Fourier transform of size $2B$ as desired. This process is shown in \Cref{fig: sharp Gabor atoms reshuffle}.

As an intermediate step of the implementation of $\snb{N, B}$, we first introduce another permutation matrix. For an integer $m$ and $M=2^m$, let $\rpm{M}$ be the $M\times M$ permutation matrix such that 
\begin{equation}
    \begin{aligned}
\rpm{M}\ket{j}  &= 
    \ket{2j},\\
\rpm{M}\ket{M-1-j} &=  \ket{2j+1},
\end{aligned}
\end{equation}
for $j\in[M/2]$.
Notice that $\rpm{M}$ can be represented by the bit manipulation
\begin{equation}
\begin{aligned}
    \ket{x_{m-1}x_{m-2}\cdots x_{0}}&\mapsto \ket{x_{m-1}(x_{m-2}\oplus x_{m-1})\cdots (x_{0}\oplus x_{m-1})}\\
    &\mapsto \ket{(x_{m-2}\oplus x_{m-1})\cdots (x_{0}\oplus x_{m-1})x_{m-1}},
\end{aligned}
\end{equation}
therefore it can be implemented on the quantum computer by $m-1$ CNOT-gates followed by $m-1$ SWAP-gates.

Now let us come back to the implementation of matrix $\snb{N,B}$. One can see that $\snb{N,B}$ only operates on the first $n-b$ qubits, and can be factorized as 
\begin{equation*}
    \snb{N,B} = \snb{\frac{N}{B},1}\otimes I_{2^b} = (I_{2^{n-b-2}}\otimes \CNOT\otimes I_{2^b})(\rpm{N/B}\otimes I_{2^b}).
\end{equation*}
Here, the extra CNOT-gate is used for deciding whether the $j$ in \eqref{eq: snb cases} is even or odd. If $j$ is odd, we need an extra switch between $\ket{2Bj+k}$ and $\ket{2Bj+B+k}$ after applying $\rpm{N/B}$ on the first $n-b$ qubits.

Finally, the complete circuit from $f$ to $a$ is given by
\begin{equation}
    \ugs = (I_{2^{n-b-1}}\otimes U_{\mathrm{FT}(2B)}^\dagger)(I_{2^{n-b-2}}\otimes \CNOT\otimes I_{2^b})(\rpm{N/B}\otimes I_{2^b}) U_{\mathrm{FT}(N)},
\end{equation}
which is demonstrated in \Cref{fig: sharp Gabor atoms}.

\begin{figure}[!ht]
\centering
\begin{small}
\begin{tikzpicture}[on grid]
\pgfkeys{/myqcircuit, layer width=8mm, row sep=5mm, source node=qwsource}
\newcommand{\qwstart}{1}
\newcommand{\qwend}{7}
\qwire[start node=qw1s, end node=qw1, style=thin, index=1, start layer=\qwstart, end layer=\qwend, label=$\ket{x_5}$]
\qwire[start node=qw2s, end node=qw2, style=thin, index=2, start layer=\qwstart, end layer=\qwend, label=$\ket{x_4}$]
\qwire[start node=qw3s, end node=qw3, style=thin, index=3, start layer=\qwstart, end layer=\qwend, label=$\ket{x_3}$]
\qwire[start node=qw1s, end node=qw4, style=thin, index=4, start layer=\qwstart, end layer=\qwend, label=$\ket{x_2}$]
\qwire[start node=qw2s, end node=qw5, style=thin, index=5, start layer=\qwstart, end layer=\qwend, label=$\ket{x_1}$]
\qwire[start node=qw3s, end node=qw6, style=thin, index=6, start layer=\qwstart, end layer=\qwend, label=$\ket{x_0}$]

\multiqubit[layer=1, start index=1, stop index=6, label=$\,U_{\mathrm{FT}(N)}\,$]
\multiqubit[layer=3, start index=1, stop index=4, label=$\,\snb{\frac{N}{B},1}\,$]
\multiqubit[layer=5, start index=4, stop index=6, label=$\,U_{\mathrm{FT}(2B)}^\dagger\,$]

\renewcommand{\qwstart}{8}
\renewcommand{\qwend}{19.5}
\qwire[start node=qw1s, end node=qw1e, style=thin, index=1, start layer=\qwstart, end layer=\qwend]
\qwire[start node=qw2s, end node=qw2e, style=thin, index=2, start layer=\qwstart, end layer=\qwend]
\qwire[start node=qw3s, end node=qw3e, style=thin, index=3, start layer=\qwstart, end layer=\qwend]
\qwire[start node=qw1s, end node=qw4e, style=thin, index=4, start layer=\qwstart, end layer=\qwend]
\qwire[start node=qw2s, end node=qw5e, style=thin, index=5, start layer=\qwstart, end layer=\qwend]
\qwire[start node=qw3s, end node=qw6e, style=thin, index=6, start layer=\qwstart, end layer=\qwend]

\multiqubit[layer=7.5+0.5, start index=1, stop index=6, label=$\,U_{\mathrm{FT}(N)}\,$]
\singlequbit[style=not, layer=7.5+2, index=2, node=not2]
\control[layer=7.5+2, index=1, target node=not2, style=controlon]
\singlequbit[style=not, layer=7.5+3, index=3, node=not3]
\control[layer=7.5+3, index=1, target node=not3, style=controlon]
\singlequbit[style=not, layer=7.5+4, index=4, node=not4]
\control[layer=7.5+4, index=1, target node=not4, style=controlon]
\swapgate[layer=7.5+5, first index=1, second index=2]
\swapgate[layer=7.5+6, first index=2, second index=3]
\swapgate[layer=7.5+7, first index=3, second index=4]
\singlequbit[style=not, layer=7.5+8, index=4, node=not42]
\control[layer=7.5+8, index=3, target node=not42, style=controlon]
\multiqubit[layer=7.5+9.5, start index=4, stop index=6, label=$\,U_{\mathrm{FT}(2B)}^\dagger\,$]
\node at ($0.5*(qw3)+0.5*(qw4)+(0.5,0)$) {$=$};
\end{tikzpicture}%
\end{small}%
\caption{The circuit of $\ugs$ when $N=64$ and $B=4$.}
\label{fig: sharp Gabor atoms}
\end{figure}

\subsection{Blended Gabor atoms}\label{sec: blended Gabor atoms}
\begin{figure}[!ht]
	\centering
	\includegraphics[
	width =0.7\textwidth
	]{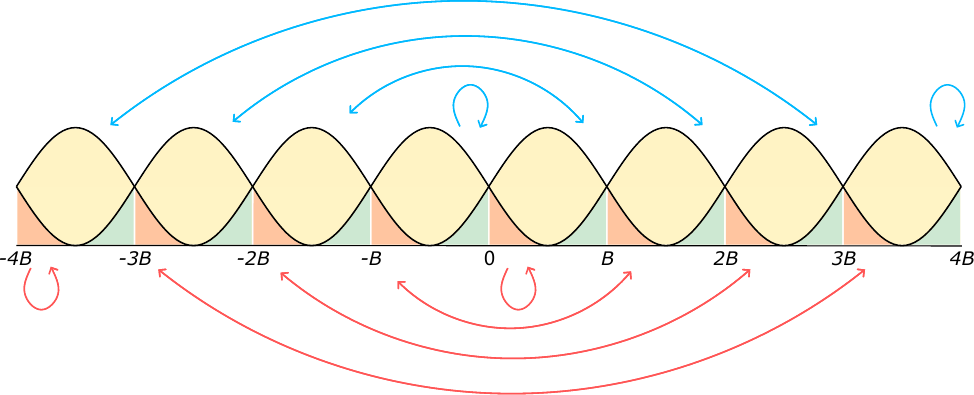}
	\caption{Illustration of reallocation when $N = 8B$. Each basis function $\fpjp$ contains two  bumps supported at $[(j-\half)B,(j+\frac{3}{2})B)$ and $[-(j+\frac{3}{2})B, -(j-\half)B)$. The arrows indicate how the small tails are transposed such that all the relevant $\hat{f}(k)$'s are placed in $[jB,(j+1)B)\cup[-(j+1)B, -jB)$ with correct proportion. All these reallocations are unitary manipulations. This figure exclusively illustrates magnitude transpositions, omitting the change of phase. In particular, the self-transpositions are, in fact, phase rotations.}
	\label{fig: packet reallocate}
\end{figure}

As discussed in \Cref{sec: background}, the aim of introducing the blended Gabor atoms is to incorporate blended frequency windows, resulting in rapid decay in the space domain. Typically, a blended Gabor atom basis function consists of two frequency bumps symmetrically positioned about the origin, such as the second and seventh bump in \Cref{fig: packet reallocate}. 

To define the discrete blended Gabor atoms, we restrict $\omega$ onto a grid and represent the index $(j,p)$ as $2Bj+p$ for convenience in ordering. The Fourier transform of a discrete blended Gabor atom is defined by
\begin{equation}\label{eq: defn wave packet}
    \fpjp(k) = \frac{1}{\sqrt{2B}}e^{2\pi i \frac{pk}{2B}}\left[ e^{\half\pi i\left(\half-\frac{k-Bj}{B}\right)} g_{\mathrm{per}}\left(\pi\left(\frac{k-Bj}{B}-\half\right)\right) + e^{\half\pi i\left(-\half-\frac{k+Bj}{B}\right)} g_{\mathrm{per}}\left(\pi\left(\frac{k+Bj}{B}+\half\right)\right)\right]
\end{equation}
for $j \in [\frac{N}{2B}]$, $p \in [2B]$, 
where
\begin{equation}
    g_{\mathrm{per}}(x) = \sum_{q\in\ZZ} g(x+q\pi\frac{N}{B})
\end{equation}
is the periodic version of $g$ defined in \eqref{eq: gx}. Here, $g_{\mathrm{per}}$ is used instead of $g$ due to the considerations of boundary conditions. This periodization also automatically makes $\fpjp(k) = \fpjp(N+k)$ hold for any integer $k$, which is compatible with our convention. 

We also point out that, akin to the sharp Gabor atom, this discrete basis function corresponds to its continuous counterpart through $\widehat{\psi_{j,p}}(\frac{2\pi}{N} k)= \sqrt{\frac{2\pi}{N}}\widehat{\uppsi_{j,p}}(k)$, except for those $k$ values near the boundary.

One can make straightforward calculations to show that $\psi_0,\ldots,\psi_{N-1}$ form an orthonormal basis of $\CC^N$. Similar to the sharp case, the coefficients of a vector $f$ are $\cpjp = \inner{\hat{f},\fpjp}$, and the blended Gabor atom transform is defined as
\begin{equation}
    \ugb:\CC^N\to \CC^N: f\mapsto a = (a_{0},a_{1},\ldots,a_{N-1})^T.
\end{equation}

In this scenario, the support of different basis functions may intersect. Therefore, we first reallocate $\hat{f}$ to the correct position. Subsequently, we perform the rearrangement of indices, identical to the procedure for sharp Gabor atoms. This reallocation process is shown in \Cref{fig: packet reallocate}.
To make it mathematically concrete, we introduce the intermediate variable $h(k)$ defined as
\begin{equation}\label{eq: defvg1}
    h(jB+q) = \hat{f}(jB+q)g\left(-\halfpi+q\frac{\pi}{B}\right)e^{\halfi(-\hlfpi+q\frac{\pi}{B})} + \hat{f}(-jB+q)g\left(\halfpi+q\frac{\pi}{B}\right)e^{\halfi(\hlfpi+q\frac{\pi}{B})},
\end{equation}
\begin{equation}\label{eq: defvg2}
    h((j+\half)B+q) = \hat{f}((j+\half)B+q)g\left(q\frac{\pi}{B}\right)e^{\halfi(q\frac{\pi}{B})} + \hat{f}(-(j+\frac{3}{2})B+q)g\left(-\pi+q\frac{\pi}{B}\right)e^{\halfi(-\pi+q\frac{\pi}{B})}
\end{equation}
for $j = -\frac{N}{2B},-\frac{N}{2B}+1,\ldots,\frac{N}{2B}-1$, $q\in [\frac{B}{2}]$. This reallocation process guarantees that
\begin{equation}\label{eq: Meyer packet h}
    \left(\sum_{k=jB}^{(j+1)B-1}+\sum_{k=-(j+1)B}^{-jB-1}\right)\frac{1}{\sqrt{2B}}e^{-2\pi i \frac{pk}{2B}}h(k) = \inner{\hat{f},\fpjp} = \cpjp. 
\end{equation}
The proof is provided in \Cref{sec: proof Meyer packet h}. Despite its complex appearance, the formulas for $h(k)$ are derived in reverse to satisfy \eqref{eq: Meyer packet h}, so the proof is straightforward calculations.

Notice that \eqref{eq: Meyer packet h} and \eqref{eq: sharp Gabor transform calculation} share the same form, so we can reuse the circuit in \Cref{sec: sharp Gabor atoms}, and conclude that the complete circuit from blended Gabor atom transform $\ugb: f\mapsto a$ is given by
\begin{equation}\label{eq: upm}
    \ugb = (I_{2^{n-b-1}}\otimes U_{\mathrm{FT}(2B)}^\dagger)\snb{N,B}\tg U_{\mathrm{FT}(N)},
\end{equation}
where $\tg$ is the transition matrix from $\hat{f}$ to $h$, and can be written down following the definitions in \eqref{eq: defvg1} and \eqref{eq: defvg2}. 

\begin{figure}[!htb]
    \centering
    \subfloat[\label{fig: block diag}]{
    \begin{minipage}[c]{0.49\textwidth}
        \centering
        \includegraphics[width=\textwidth]{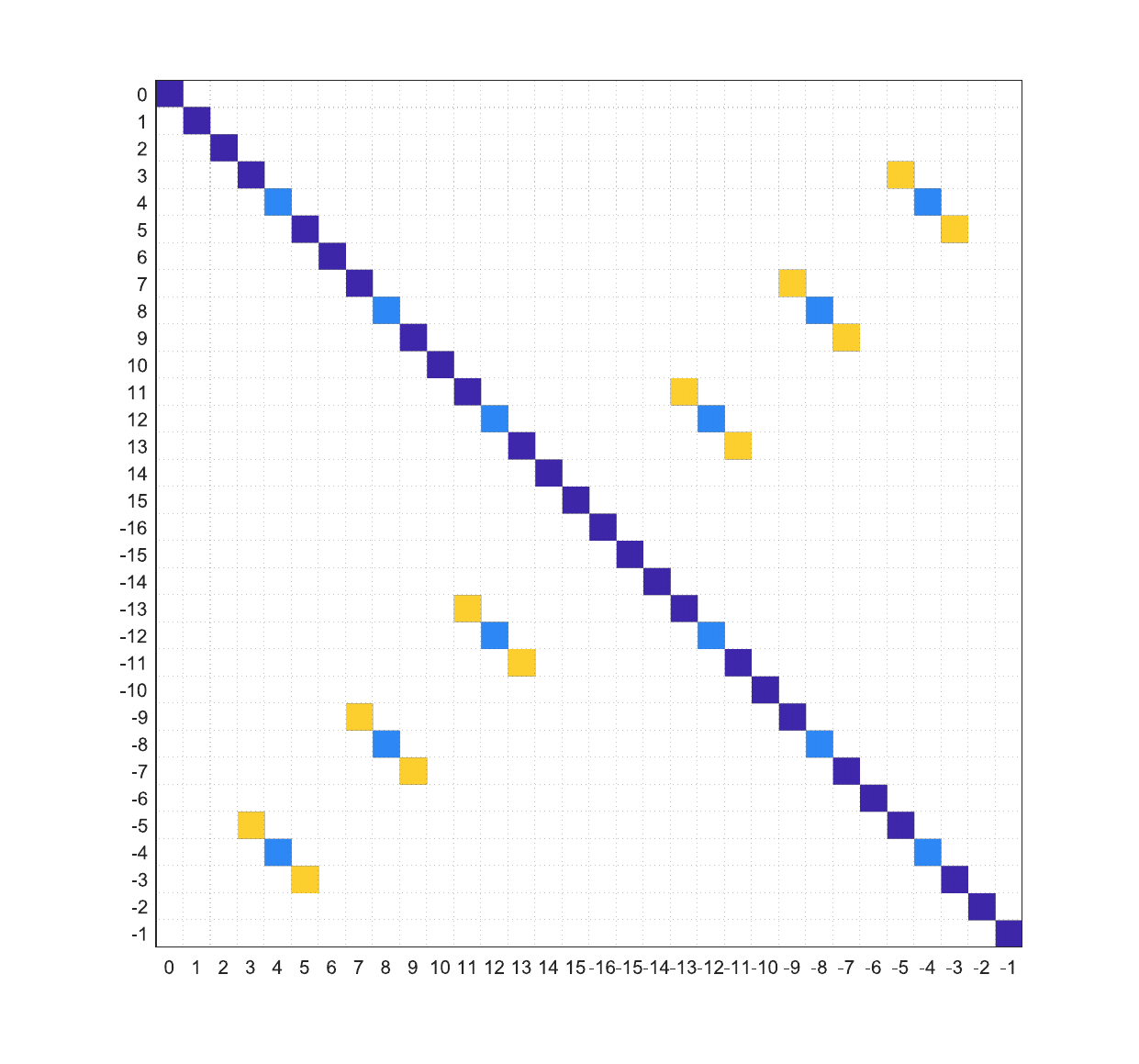}
    \end{minipage}
    }
    \subfloat[\label{fig: vg}]{
    \begin{minipage}[c]{0.49\textwidth}
        \centering
        \includegraphics[width=\textwidth]{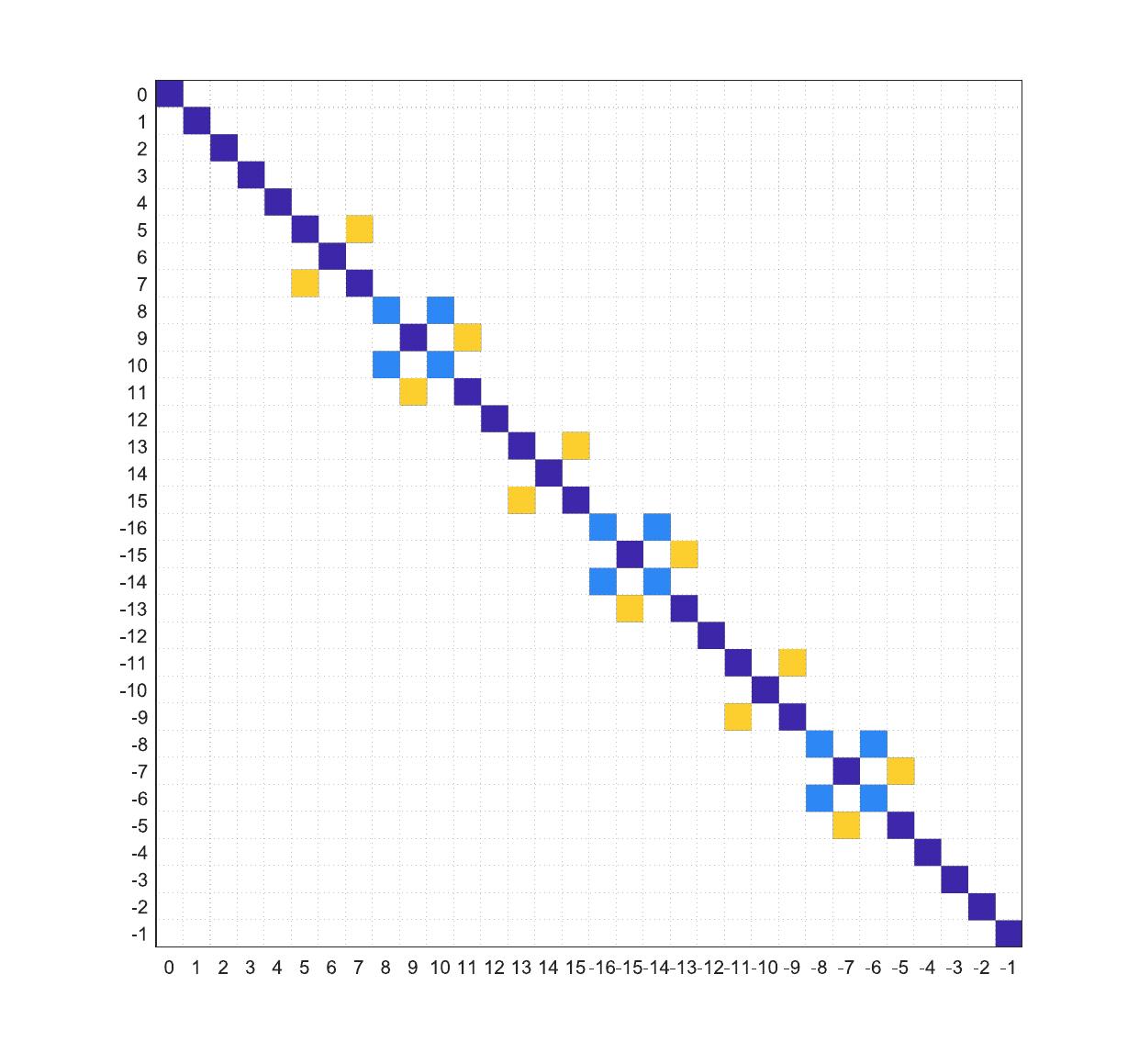}
    \end{minipage}
    }
    \caption{(a) Magnitude of elements in $\tg$ with $N = 32$, $B=4$. (b) Magnitude of elements in $\vg$ with $N = 32$, $B=4$.}
    \label{fig: block diag structure}
\end{figure}

Let us now focus on the implementation of $\tg$. Since $\tg$ has the potential block-diagonal structure, as shown in \Cref{fig: block diag}, we shall first rearrange the order of the computational basis to reveal this structure. At this stage, we do not need to deal with the last $b-1$ qubits. For an integer $M = 2^m$, we may define the permutation matrix $\qpm{M}$ as
\begin{equation}
\begin{aligned}
    \qpm{M}\ket{j} &= \ket{2j}, && \text{for $j\in [M/2]$,}\\
    \qpm{M}\ket{M-1} &= \ket{M-1},\\
    \qpm{M}\ket{M/2} &= \ket{1},\\
    \qpm{M}\ket{M-2j} &= \ket{4j+1}, && \text{for $j = 1,2,\ldots, M/4-1$,}\\
    \qpm{M}\ket{M-2j-1} &= \ket{4j-1}, && \text{for $j = 1,2,\ldots, M/4-1$.}
\end{aligned}    
\end{equation}
The detailed implementation of $\qpm{M}$ is shown in \Cref{sec: implement of qpm}. We can use the permutation matrix $\qpm{\frac{2N}{B}}\otimes I_{2^{b-1}}$ to rearrange the nonzero elements in $\tg$ and define a new matrix
\begin{equation}\label{eq: tg d}
    \vg = (\qpm{\frac{2N}{B}}\otimes I_{2^{b-1}})^\dagger \tg (\qpm{\frac{2N}{B}}\otimes I_{2^{b-1}}),
\end{equation}
which is a block-diagonal matrix as shown in \Cref{fig: vg}. A detailed description of $\vg$ and its implementation is discussed in \Cref{sec: implementation of vg}. Combining \eqref{eq: upm} and \eqref{eq: tg d}, we can draw the complete circuit of $\ugb$ in \Cref{fig: blended Gabor atoms}.

\begin{figure}[!ht]
\centering
\begin{small}
\begin{tikzpicture}[on grid]
\pgfkeys{/myqcircuit, layer width=8mm, row sep=5mm, source node=qwsource}
\newcommand{\qwstart}{1}
\newcommand{\qwend}{12.5}
\qwire[start node=qw1s, end node=qw1, style=thin, index=1, start layer=\qwstart, end layer=\qwend, label=$\ket{x_5}$]
\qwire[start node=qw2s, end node=qw2, style=thin, index=2, start layer=\qwstart, end layer=\qwend, label=$\ket{x_4}$]
\qwire[start node=qw3s, end node=qw3, style=thin, index=3, start layer=\qwstart, end layer=\qwend, label=$\ket{x_3}$]
\qwire[start node=qw1s, end node=qw4, style=thin, index=4, start layer=\qwstart, end layer=\qwend, label=$\ket{x_2}$]
\qwire[start node=qw2s, end node=qw5, style=thin, index=5, start layer=\qwstart, end layer=\qwend, label=$\ket{x_1}$]
\qwire[start node=qw3s, end node=qw6, style=thin, index=6, start layer=\qwstart, end layer=\qwend, label=$\ket{x_0}$]

\multiqubit[layer=1, start index=1, stop index=6, label=$\,U_{\mathrm{FT}(N)}\,$]
\multiqubit[layer=3, start index=1, stop index=5, label=$\,\qpm{\frac{2N}{B}}^\dagger\,$]
\multiqubit[layer=4.5, start index=1, stop index=6, label=$\,\vg\,$]
\multiqubit[layer=6, start index=1, stop index=5, label=$\,\qpm{\frac{2N}{B}}\,$]
\multiqubit[layer=8, start index=1, stop index=4, label=$\,\snb{\frac{N}{B},1}\,$]
\multiqubit[layer=10, start index=4, stop index=6, label=$\,U_{\mathrm{FT}(2B)}^\dagger\,$]
\end{tikzpicture}%
\end{small}%
\caption{The circuit of $\ugb$ when $N=64$ and $B=4$.}
\label{fig: blended Gabor atoms}
\end{figure}

\subsubsection{The implementation of $\qpm{M}$}\label{sec: implement of qpm}

Let $L$ be the $M\times M$ shift matrix
\begin{equation}
    L =
\begin{bmatrix}
0      & 0      & \hdots & \hdots & 1\\
1      & 0      & 0      & \ddots & 0 \\
\vdots & 1      & \ddots & \ddots & \vdots \\
\vdots & \ddots & \ddots & \ddots & \vdots \\
0      & 0      & \hdots & 1      & 0 \\
\end{bmatrix},
\end{equation}
which can be implemented easily within $O(m^2)$ gate complexity, as shown in \cite[Figure 6]{camps2022explicit}. If one is allowed to use one ancilla qubit, then $L$ can be implemented within $O(m)$ gate complexity as in \cite{klappenecker1999wavelets}.

\begin{figure}[!ht]
\centering
\begin{small}
\begin{tikzpicture}[on grid]
\pgfkeys{/myqcircuit, layer width=8mm, row sep=5mm, source node=qwsource}
\newcommand{\qwstart}{1}
\newcommand{\qwend}{7}
\qwire[start node=qw1s, end node=qw1e, style=thin, index=1, start layer=\qwstart, end layer=\qwend, label=$~$]
\qwire[start node=qw2s, end node=qw2e, style=thin, index=2, start layer=\qwstart, end layer=\qwend, label=$~$]
\qwire[start node=qw3s, end node=qw3e, style=thin, index=3, start layer=\qwstart, end layer=\qwend, label=$~$]
\qwire[start node=qw1s, end node=qw1e, style=thin, index=4, start layer=\qwstart, end layer=\qwend, label=$~$]
\qwire[start node=qw2s, end node=qw2e, style=thin, index=5, start layer=\qwstart, end layer=\qwend, label=$~$]

\multiqubit[layer=1, start index=2, stop index=5, label=$L^\dagger$]
\singlequbit[style=not, layer=2, index=5, node=x1]
\control[layer=2, index=1, target node=x1, style=controlon]
\multiqubit[layer=3, start index=2, stop index=5, label=$L$]
\multiqubit[layer=4.5, start index=1, stop index=5, label=$\,\rpm{M}\,$]

\end{tikzpicture}%
\end{small}%
\caption{The circuit of $\qpm{M}$ for $M = 32$.}
\label{fig: pm}
\end{figure}
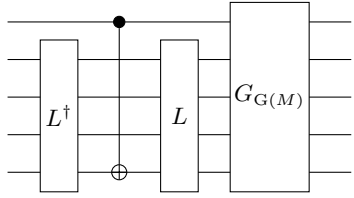

Notice that $\qpm{M}$ has the same $M/2$ columns as the matrix $\rpm{M}$ introduced earlier, so it is natural to reuse the circuit of $\rpm{M}$ in the construction of $\qpm{M}$. Straightforward calculations show that $\rpm{M}^{-1}\qpm{M}$ is a permutation matrix that switches $\ket{M-2j}$ and $\ket{M-2j-1}$ for $j = 1,2,\ldots,\frac{M}{4}-1$, and also switches $\ket{M-1}$ and $\ket{M/2}$. Therefore, we have $\rpm{M}^{-1}\qpm{M} = \ketbra{0}{0}\otimes I_{M/2}+\ketbra{1}{1}\otimes(L(I_{M/4}\otimes X)L^\dagger)$. Finally, noticing that the controls on $L$ and $L^\dagger$ are unnecessary since they will be canceled, we conclude
\begin{equation}
    \qpm{M} = \rpm{M} (I_2\otimes L) \left(\ketbra{0}{0}\otimes I_{M/2}+\ketbra{1}{1}\otimes I_{M/4}\otimes X \right) (I_2\otimes L)^\dagger,
\end{equation}
where $\left(\ketbra{0}{0}\otimes I_{M/2}+\ketbra{1}{1}\otimes I_{M/4}\otimes X \right)$ is just the CNOT gate on the first and the last qubits. \Cref{fig: pm} illustrates the circuit implementation of $\qpm{M}$.

\subsubsection{Implementation of $\vg$}\label{sec: implementation of vg}
To help the following discussion, we define two diagonal matrices $D_+ = \diag\{0,\frac{1}{B},\frac{2}{B},\ldots,\frac{(B/2-1)}{B}\}$ and $D_- = D_+ - \half I$.

As illustrated in \Cref{fig: vg}, the matrix $\vg$ is a block-diagonal matrix with $\frac{N}{B}$ blocks, with each block of size $B\times B$. By the definitions in \eqref{eq: defvg1} and \eqref{eq: defvg2} together with the permutation matrix $\qpm{M}$, we can write down the blocks explicitly. 
\begin{equation}\label{eq: blocks of qpm}
\begin{aligned}
    \text{The 0-th block } = \singleeven &:= I_2\otimes
        e^{\halfi(\pi\beta(D_-)+\pi D_-)}.\\
    \text{The $(2j)$-th block } = \twoeven &:= \begin{bmatrix}
        \cos(\halfpi\beta(D_-))e^{\halfi \pi D_-} & i\sin(\halfpi\beta(D_-))e^{\halfi \pi D_-}\\
        i\sin(\halfpi\beta(D_-))e^{\halfi \pi D_-} & \cos(\halfpi\beta(D_-))e^{\halfi \pi D_-} 
    \end{bmatrix} \\
    &= (H\otimes I_{B/2}) \begin{bmatrix}
        e^{\halfi(\pi\beta(D_-)+\pi D_-)} & 0\\
        0 & e^{\halfi(-\pi\beta(D_-)+\pi D_-)}
    \end{bmatrix}(H\otimes I_{B/2}), \\
    &\text{for $j =1,2,\ldots, \frac{N}{2B}-1$.}\\
    \text{The $(2j-1)$-th block } = \twoodd &:= \begin{bmatrix}
        \cos(\halfpi\beta(D_+))e^{\halfi \pi D_+} & -i \sin(\halfpi\beta(D_+))e^{\halfi \pi D_+}\\
        -i\sin(\halfpi\beta(D_+))e^{\halfi \pi D_+} & \cos(\halfpi\beta(D_+))e^{\halfi \pi D_+} 
    \end{bmatrix} \\
    &= (H\otimes I_{B/2}) \begin{bmatrix}
        e^{\halfi(-\pi\beta(D_+)+\pi D_+)} & 0\\
        0 & e^{\halfi(\pi\beta(D_+)+\pi D_+)}
    \end{bmatrix}(H\otimes I_{B/2}), \\
    &\text{for $j =1,2,\ldots, \frac{N}{2B}-1$.}\\
    \text{The $(\frac{N}{B}-1)$-th block } = \singleodd &:= I_2\otimes
        e^{\halfi(-\pi\beta(D_+)+\pi D_+)}.\\
\end{aligned}
\end{equation}
The remaining challenge lies in implementing these diagonal matrices. Usually, $\beta(x)$ will be chosen as a polynomial on $[0,\half]$ and then extend to $[-1,1]$ according to \eqref{eq: constraints on beta}. Typical choices include $\beta(x) = x$ or $\beta(x)=2x^2$ in the interval $[0,\half]$. For such low-degree $\beta$, one can use the method introduced in \Cref{sec: exp i poly A bit} to implement the required diagonal matrices. If one wants to pursue higher regularity with more sophisticated $\beta(x)$, such as 
\begin{equation}\label{eq: 7 deg beta}
    \beta(x) = x^4(35-84x+70x^2-20x^3),\quad \text{for } x\in [0,1],
\end{equation}
or even non-polynomial functions, the method in \Cref{sec: exp i poly A qsvt} could be used.

Finally, the circuit of $\vg$ is presented in \Cref{fig: circuit of vg}. Here, we need to implement the multi-qubit controlled version of a unitary. A naive way is to do the multi-qubit controlled version of every gate in this unitary in order. However, this approach results in very high gate complexity. Instead, we may introduce an ancilla qubit to make the implementation using mainly single qubit controls, as demonstrated in \Cref{fig: multi-qubit control}.

\begin{figure}[!ht]
\centering
\begin{small}
\begin{tikzpicture}[on grid]
\pgfkeys{/myqcircuit, layer width=8mm, row sep=5mm, source node=qwsource}
\newcommand{\qwstart}{1}
\newcommand{\qwend}{8}
\qwire[start node=qw1s, end node=qw1, style=thin, index=1, start layer=\qwstart, end layer=\qwend]
\qwire[start node=qw2s, end node=qw2, style=thin, index=2, start layer=\qwstart, end layer=\qwend]
\qwire[start node=qw3s, end node=qw3, style=thin, index=3, start layer=\qwstart, end layer=\qwend]
\qwire[start node=qw1s, end node=qw4, style=thin, index=4, start layer=\qwstart, end layer=\qwend]
\qwire[start node=qw2s, end node=qw5, style=thin, index=5, start layer=\qwstart, end layer=\qwend]
\qwire[start node=qw3s, end node=qw6, style=thin, index=6, start layer=\qwstart, end layer=\qwend]

\multiqubit[layer=1, start index=5, stop index=6, label=$\,\twoeven\,$, node = Y]
\control[layer=1, index=1, target node=Y, style=controloff, node = Z]

\multiqubit[layer=2.5, start index=5, stop index=6, label=$\,\twoeven^\dagger\singleeven\,$, node = Y]
\control[layer=2.5, index=4, target node=Y, style=controloff, node = Y1]
\control[layer=2.5, index=3, target node=Y1, style=controloff, node = Y2]
\control[layer=2.5, index=2, target node=Y2, style=controloff, node = Y3]
\control[layer=2.5, index=1, target node=Y3, style=controloff, node = Y4]

\multiqubit[layer=4, start index=5, stop index=6, label=$\,\twoodd\,$, node = Y]
\control[layer=4, index=1, target node=Y, style=controlon, node = Z]

\multiqubit[layer=5.5, start index=5, stop index=6, label=$\,\twoodd^\dagger\singleodd\,$, node = Y]
\control[layer=5.5, index=4, target node=Y, style=controlon, node = Y1]
\control[layer=5.5, index=3, target node=Y1, style=controlon, node = Y2]
\control[layer=5.5, index=2, target node=Y2, style=controlon, node = Y3]
\control[layer=5.5, index=1, target node=Y3, style=controlon, node = Y4]

\end{tikzpicture}%
\end{small}%
\caption{The circuit of $\vg$ when $N=64$ and $B=4$.}
\label{fig: circuit of vg}
\end{figure}

\begin{figure}[!ht]
\centering
\begin{small}
\begin{tikzpicture}[on grid]
\pgfkeys{/myqcircuit, layer width=8mm, row sep=5mm, source node=qwsource}
\newcommand{\qwstart}{1}
\newcommand{\qwend}{3}
\qwire[start node=qw2s, end node=qw2, style=thin, index=2, start layer=\qwstart, end layer=\qwend]
\qwire[start node=qw3s, end node=qw3, style=thin, index=3, start layer=\qwstart, end layer=\qwend]
\qwire[start node=qw1s, end node=qw4, style=thin, index=4, start layer=\qwstart, end layer=\qwend]
\qwire[start node=qw2s, end node=qw5, style=thin, index=5, start layer=\qwstart, end layer=\qwend]
\qwire[start node=qw3s, end node=qw6, style=thin, index=6, start layer=\qwstart, end layer=\qwend]

\multiqubit[layer=1, start index=5, stop index=6, label=$\,U\,$, node = Y]
\control[layer=1, index=4, target node=Y, style=controlon, node = Y1]
\control[layer=1, index=3, target node=Y1, style=controlon, node = Y2]
\control[layer=1, index=2, target node=Y2, style=controlon, node = Y3]

\renewcommand{\qwstart}{6}
\renewcommand{\qwend}{10}
\qwire[start node=qw1s, end node=qw1e, style=thin, index=1, start layer=\qwstart, end layer=\qwend, label = ancilla $\ket{0}$]
\qwire[start node=qw2s, end node=qw2e, style=thin, index=2, start layer=\qwstart, end layer=\qwend]
\qwire[start node=qw3s, end node=qw3e, style=thin, index=3, start layer=\qwstart, end layer=\qwend]
\qwire[start node=qw1s, end node=qw4e, style=thin, index=4, start layer=\qwstart, end layer=\qwend]
\qwire[start node=qw2s, end node=qw5e, style=thin, index=5, start layer=\qwstart, end layer=\qwend]
\qwire[start node=qw3s, end node=qw6e, style=thin, index=6, start layer=\qwstart, end layer=\qwend]

\singlequbit[style=not, layer=6, index=1 , node = Y]
\control[layer=6, index=2, target node=Y, style=controlon, node = Y1]
\control[layer=6, index=3, target node=Y1, style=controlon, node = Y2]
\control[layer=6, index=4, target node=Y2, style=controlon, node = Y3]

\multiqubit[layer=7, start index=5, stop index=6, label=$\,U\,$, node = Y]
\control[layer=7, index=1, target node=Y, style=controlon, node = Y1]

\singlequbit[style=not, layer=8, index=1 , node = Y]
\control[layer=8, index=2, target node=Y, style=controlon, node = Y1]
\control[layer=8, index=3, target node=Y1, style=controlon, node = Y2]
\control[layer=8, index=4, target node=Y2, style=controlon, node = Y3]

\node at ($0.5*(qw3)+0.5*(qw4)+(0.5,0)$) {$=$};

\pgfkeys{/myqcircuit, gate offset=1}
\renewcommand{\qwstart}{11}
\renewcommand{\qwend}{11}

\qwire[index=1, start layer=\qwstart, end layer=\qwend, label = $\ket{0}$]

\end{tikzpicture}%
\end{small}%
\caption{The circuit of implementing multi-qubit controlled unitaries using one ancilla qubit.}
\label{fig: multi-qubit control}
\end{figure}
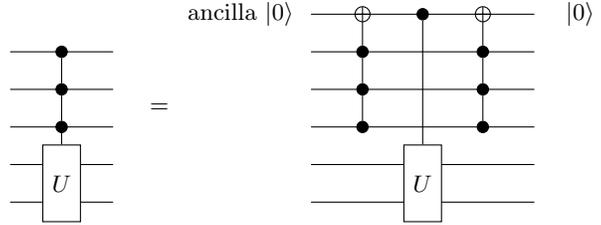

\subsubsection{Implementation of the exponential of a pure imaginary polynomial via bit manipulations}\label{sec: exp i poly A bit}
Now we focus on the general problem of implementing diagonal matrices $\exp (i q(A))$, where $A = \diag\{0,1,\ldots,M-1\}$ for some $M = 2^m$, and $q$ is a real polynomial. 
Since polynomials are the sum of monomials, we only need to consider the special case $q(A) = rA^s$, where $r\in\RR$ and $s\ge 0$ is integer. The case $s=0$ is trivial. The case $s=1$ is discussed in \cite[Section 5.1]{li2023efficient}, which exploits the following observation. Let $x\in [M]$ and $x = (x_{m-1}\cdots x_0)_2$ be its binary representation, then
\begin{equation}
    \exp (irA)\ket{x} =\exp (irx)\ket{x} = \bigotimes_{j = 0}^{m-1} \left(\exp(ir x_j 2^{j})\ket{x_j}\right) = \bigotimes_{j = 0}^{m-1} \left(\rz(2^j r)\ket{x_j}\right).
\end{equation}
Where $\rz(\theta) := \begin{pmatrix}
    1&\\&e^{i\theta}
\end{pmatrix}$ is the single-qubit rotation gate.
This means $\exp(irA)$ can be implemented by simply applying an $\rz(2^j r)$ on the $j$-th qubit.

For general $s$, a similar approach can be utilized, but it may involve multi-qubit gates due to the cross terms. Specifically, we have
\begin{equation}
\begin{aligned}
    x^s = \left(\sum_{j=0}^{m-1}2^j x_j\right)^s &= \sum_{k_0+k_1+\cdots+k_{m-1}=s ;\, k_j\geq 0}\left(\begin{array}{c}
n \\
k_0, k_1, \ldots, k_{m-1}
\end{array}\right) \prod_{j=0}^{m-1}(2^{jk_j} x_j^{k_j})\\
&= \sum_{k_0+k_1+\cdots+k_{m-1}=s ;\, k_j\geq 0}\left(\begin{array}{c}
n \\
k_0, k_1, \ldots, k_{m-1}
\end{array}\right) (\prod_{j=0}^{m-1}2^{jk_j})\prod_{k_j>0} x_j.
\end{aligned}
\end{equation}
Therefore, to implement $\exp (irx^s)$, we only need to implement the terms look like $\exp (ic\prod_{j\in J}x_j)$, where $J$ is an index set, and $c$ is a real number. This $2^{|J|}$ dimensional matrix can be realized as a controlled version of $\rz(c)$, which is \begin{equation}
    \diag\{1,1,\ldots,1,e^{ic}\} = \ket{1^{|J|-1}}\bra{1^{|J|-1}}\otimes \rz(c) + (I-\ket{1^{|J|-1}}\bra{1^{|J|-1}})\otimes I.
\end{equation} Note that it does not matter which $|J|-1$ qubits serve as control qubits since this matrix will not change under different choices.

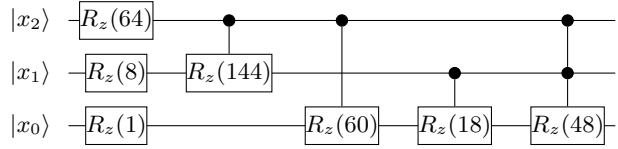
\begin{figure}[!ht]
\centering
\begin{small}
\begin{tikzpicture}[on grid]
\pgfkeys{/myqcircuit, layer width=7.5mm, row sep=7mm, source node=qwsource}
\newcommand{\qwstart}{1}
\newcommand{\qwend}{11}
\qwire[index=1, start layer=\qwstart, end layer=\qwend, label=$\ket{x_2}$]
\qwire[index=2, start layer=\qwstart, end layer=\qwend, label=$\ket{x_1}$, end node=qw2a]
\qwire[index=3, start layer=\qwstart, end layer=\qwend, label=$\ket{x_0}$, end node=qw3a]

\singlequbit[style=gate, layer=-1+2*1, index=1, node=H1, label=$\rz(64)$]
\singlequbit[style=gate, layer=-1+2*1, index=2, node=H1, label=$\rz(8)$]
\singlequbit[style=gate, layer=-1+2*1, index=3, node=H1, label=$\rz(1)$]
\singlequbit[style=gate, layer=-1+2*2, index=2, node=H2, label=$\rz(144)$]
\control[layer=-1+2*2, index=1, target node=H2, style=controlon]
\singlequbit[style=gate, layer=-1+2*3, index=3, node=H3, label=$\rz(60)$]
\control[layer=-1+2*3, index=1, target node=H3, style=controlon]
\singlequbit[style=gate, layer=-1+2*4, index=3, node=H4, label=$\rz(18)$]
\control[layer=-1+2*4, index=2, target node=H4, style=controlon]
\singlequbit[style=gate, layer=-1+2*5, index=3, node=H5, label=$\rz(48)$]
\control[layer=-1+2*5, index=2, target node=H5, node = ctrl1, style=controlon]
\control[layer=-1+2*5, index=1, target node=ctrl1, style=controlon]

\end{tikzpicture}%
\end{small}%
\caption{The implementation of the unitary $\sum_{x=0}^{7}\exp(ix^3)\ketbra{x}{x}$.}
\label{fig: cubic}
\end{figure}

To demonstrate this clearly, we give an example for $m=3$, $s=3$, $r=1$. In this case,
\begin{equation}
    \exp (ix^3)\ket{x} = \exp(i(x_0+8x_1+64x_2+18x_0x_1+60x_0x_2+144x_1x_2+48x_0x_1x_2))\ket{x_2x_1x_0},
\end{equation}
which can be implemented as \Cref{fig: cubic}.

This method's advantage is that it does not use any ancilla qubit and does not introduce any approximation, so it is desirable as long as we have a moderate value of $s$. 

\subsubsection{Implementation of the exponential of a pure imaginary polynomial via QSVT}\label{sec: exp i poly A qsvt}
The complexity of the implementation given in \Cref{sec: exp i poly A bit} is $O(m^s)$, where $m$ is the number of qubits, and $s$ is the degree of polynomial $q$. This complexity is unaffordable when the degree $s$ is large.  

For polynomial $q$ of higher degrees, one may resort to QSVT \cite{GilyenSuLowEtAl2018} for better asymptotic complexity scaling at the expense of introducing approximations and extra ancilla qubits. To be specific, we shall construct a $(\sqrt{2},3,\epsilon)$-block encoding of the desired unitary matrix using QSVT, where $\epsilon$ is the tolerable error, and then use the ``perfect amplitude amplification'' \cite{Bagherimehrab2023efficient} to extract the desired matrix. (One may refer to \Cref{sec: notation} for the terminology regarding block encoding.)

Recall that $\beta$ is a polynomial on $[-1,1]$ satisfying \eqref{eq: constraints on beta}, and a typical example is \eqref{eq: 7 deg beta}. Therefore, $\gamma(x):= 2\beta(x+\half)-1$ is an odd polynomial. 
The diagonal matrices that we need to implement in \eqref{eq: blocks of qpm} are $e^{\halfi(\pi\beta(D_-)+\pi D_-)}$, $e^{\halfi(-\pi\beta(D_-)+\pi D_-)}$, $e^{\halfi(-\pi\beta(D_+)+\pi D_+)}$, and $e^{\halfi(\pi\beta(D_+)+\pi D_+)}$.

Now we consider $e^{\halfi(\pi\beta(D_-)+\pi D_-)}$ as an example since the implementations for the other three matrices are analogous. In this case, the number of qubits is $m=b-1=\log_2B-1$. Noticing that
$$e^{\halfi \pi(\beta(D_-)+D_-)} = e^{\halfi \pi(-\beta(-D_-)+D_-)} = e^{\halfi \pi(-\half(\gamma(-D_--\half)+1)+D_-)} = e^{-\frac{i\pi}{4}\gamma(-D_--\half)}e^{-\frac{i\pi}{4}+\frac{i\pi}{2}D_-},$$
where $e^{-\frac{i\pi}{4}+\frac{i\pi}{2}D_-}$ can be easily implemented using the method in \Cref{sec: exp i poly A bit}, we focus on 
$$e^{-\frac{i\pi}{4}\gamma(W)} = \cos(\frac{\pi}{4}\gamma(W)) -i\sin(\frac{\pi}{4}\gamma(W)),$$
where $W := -D_--\half$ for the simplicity of notations.
We need to first implement the $(1,2,\epsilon)$-block encodings of $\cos(\frac{\pi}{4}\gamma(W))$ and $\sin(\frac{\pi}{4}\gamma(W))$, and then use LCU \cite[Section 7.3]{Lin_qasc} to get a $(\sqrt{2},3,\epsilon)$-block encoding of $e^{-\frac{i\pi}{4}\gamma(W)}$. Below, we explain this idea in more detail.

For the $(1,2,\epsilon)$-block encoding of $\cos(\frac{\pi}{4}\gamma(W))$, we outline the procedure following the methodology detailed in \cite[Section 5.1]{li2023efficient}.  First, we can use the method in \Cref{sec: exp i poly A bit} to construct the $(1,1,0)$-block encoding of $\sin(W)$, given by
$$\begin{bmatrix}
    \sin(W) & -i\cos(W)\\ -i\cos(W) & \sin(W)
\end{bmatrix} = (H\otimes I)\begin{bmatrix}
    -iI & 0\\  & iI
\end{bmatrix}\begin{bmatrix}
    e^{iW} & 0\\ 0 & e^{-iW}
\end{bmatrix}(H\otimes I).$$
As $-\half I\le W\le \half I$, we seek an $\eps$ accurate polynomial approximation of the even function $\cos(\frac{\pi}{4}(\gamma(\arcsin(x))))$ for $x\in[-\sin(\half),\sin(\half)]$ to utilize QSVT on $\sin(W)$ to obtain a $(1,2,\eps)$-block encoding of $\cos(\frac{\pi}{4}(\gamma(W)))$. Let $m_{\gamma} = \max_{-\half\le x\le \half}|\gamma'(x)|$. By truncating the Taylor series of $\arcsin(x)$ at $x=0$ to degree $O(\log(\frac{2m_{\gamma}}{\eps}))$, we get a polynomial $w(x)$ satisfying $|w(x)-\arcsin(x)|\le \frac{\eps}{2m_{\gamma}}$ for $|x|\le \sin(\half)$. Consequently, $|\gamma(w(x))-\gamma(\arcsin(x))|\le \frac{\eps}{2}$. We can also expand $\cos(\frac{\pi}{4}y)$ as Taylor series at $y=0$ and truncate at degree $O(\log\frac{1}{\eps})$ to obtain a polynomial $r(y)$ such that $|r(y)-\cos(\frac{\pi}{4}y)|\le \frac{\eps}{2}$ for $|y|\le 1$. Given that $|\gamma(x)|\le 1$ for $x\in[-\half,\half]$, we finally derive
$$\begin{aligned}
    & \abs{r(\gamma(w(x))) - \cos(\frac{\pi}{4}(\gamma(\arcsin(x))))}\\\
    \le {}& \abs{r(\gamma(w(x))) - \cos(\frac{\pi}{4}\gamma(w(x)))}+\abs{\cos(\frac{\pi}{4}\gamma(w(x))) - \cos(\frac{\pi}{4}(\gamma(\arcsin(x))))}\\
    \le {}& \frac{\eps}{2}+\frac{\eps}{2}\max_y \abs{\frac{\mathrm{d}\cos(\frac{\pi}{4}y)}{\mathrm{d}y}}\le \eps.
\end{aligned}$$
Therefore, this polynomial $r(\gamma(w(x)))$ of degee $O(\log(\frac{1}{\eps})\log(\frac{2m_{\gamma}}{\eps})\deg(\gamma))$ is an $\eps$ accurate polynomial approximation of the even function $\cos(\frac{\pi}{4}(\gamma(\arcsin(x))))$. After this polynomial is constructed, we can find the corresponding phase factors and implement QSVT, which we refer to \cite{ying2022stable, dong2022infinite, dong2023robust} for details. Typically, $\gamma$ is a smooth function of moderate degree, so we may regard $m_{\gamma}$ and $\deg{\gamma}$ as $O(1)$ and conclude that the total cost of implementing a $(1,2,\epsilon)$-block encoding of $\cos(\frac{\pi}{4}\gamma(W))$ is $O(b\log(\frac{1}{\eps})^2)$, taking the encoding cost of $\sin(W)$ into account. The $(1,2,\eps)$-block encoding of $\sin(\frac{\pi}{4}\gamma(W))$ follows the same routine, and the only difference is one shall approximate the odd function $\sin(\frac{\pi}{4}(\gamma(\arcsin(x))))$. Using LCU, we can implement a $(\sqrt{2},3,\epsilon)$-block encoding of $e^{\halfi(\pi\beta(D_-)+\pi D_-)}$ within $O(b\log(\frac{1}{\eps})^2)$ cost. When measuring the ancilla qubits, we get the desired matrix for probability $\half$. A naive way to extract the desired unitary matrix is to repeat the whole process until success. However, a better way is to use the ``perfect amplitude amplification'' \cite{Bagherimehrab2023efficient} to boost the success probability to $1$ deterministically with a constant overhead. Therefore, we conclude that we can implement $e^{\halfi(\pi\beta(D_-)+\pi D_-)}$ within $O(b\log(\frac{1}{\eps})^2) = O(n\log(\frac{1}{\eps})^2)$ cost.

\subsection{Complexity analysis}

The circuit complexity of our Gabor atom implementation is analyzed below, assuming single-qubit and two-qubit gates as elementary building blocks. The circuit of the sharp Gabor atoms, shown in \Cref{fig: sharp Gabor atoms}, is $O(n^2)$, comprising $O(n)$ operations flanked by two Fourier transform circuits, each with $O(n^2)$ complexity. For blended Gabor atoms, the complexity remains at most $O(\mathrm{poly}(n))$, even if we always use the exact implementations and do not use any ancilla qubit as in \Cref{fig: multi-qubit control}. For instance, selecting $\beta(x)$ as $x$ or $2x^2$ for $0\le x\le \frac{1}{2}$ and employing the method in \Cref{sec: exp i poly A bit} for diagonal matrix implementation still yields $O(n^2)$ complexity, since we can check that each block in \Cref{fig: blended Gabor atoms} costs no more than $O(n^2)$.

If we allow for an $\eps$ error and use three ancilla qubits, then the overall complexity is only $O(n(\log n+\log(\eps^{-1})^2))$. A step-by-step examination of \Cref{fig: blended Gabor atoms} verifies this. The Fourier transform can be performed using an $O(n\log(n\epsilon^{-1}))$ algorithm \cite{nam2020approximate}. $\qpm{M}$ has an $O(n)$ complexity since $\rpm{M}$ is $O(n)$ and $L$ can also be implemented within $O(n)$ complexity using one ancilla qubit \cite{klappenecker1999wavelets}. In the implementation of $\vg$, the multiple qubit controls can be implemented as in \Cref{fig: multi-qubit control}, and the required diagonal matrices can be implemented within $O(n\log(\eps^{-1})^2)$ for general $\beta(x)$ using three ancilla qubits as discussed in \Cref{sec: exp i poly A qsvt}.

\section{Wavelets}\label{sec: wavelet}

This section gives the discretization of the (sharp) Shannon wavelets and (blended) Meyer wavelets and presents the corresponding quantum circuits. The discretization of the Shannon wavelet is straightforward, and its circuit implementation can be viewed as a recursive process. For the Meyer wavelets, a phase adjustment is necessary to ensure orthogonality when imposing periodic boundary conditions. When implementing its quantum circuit, we first reallocate the Fourier coefficient $\hat{f}(k)$'s into the correct positions and then apply the circuit for the Shannon wavelet, akin to our approach with Gabor atoms. However, the reallocation process involves multiple levels of control, and sometimes the control occurs at $\frac{N}{3\cdot 2^j}$ rather than powers of 2, which makes it more technical.

\subsection{Shannon wavelets (sharp)}
In the continuous setting, the Shannon wavelet basis $\widehat{\uppsi_{j,p}}$ is given by \eqref{eq: wavelet cont} and \eqref{eq: Shannon wavelet cont}. The Fourier transforms of discrete Shannon wavelets are given by
\begin{equation}
    \fwjp (k) = \frac{1}{\sqrt{2^{n-j}}} e^{2\pi i\frac{pk}{2^{n-j}}}\left[ \chi_{[-2^{n-j},-2^{n-j-1})}(k) + \chi_{[2^{n-j-1},2^{n-j})}(k)\right],
\end{equation}
for $j=1,2,\ldots,n$ and $p\in [2^{n-j}]$. In particular, when $j=n$ and $p=0$, $\fwjp = \chi_{\{-1\}}$. In addition, the last basis functions are given by the scaling function
\begin{equation}
    \widehat{\phi_{n,0}} = \chi_{\{0\}}.
\end{equation}
We remark that, similar to the sharp Gabor atom case, it holds that $\widehat{\psi_{j,p}}(k)= \sqrt{\frac{2\pi}{N}}\widehat{\uppsi_{j,p}}(\frac{2\pi}{N} k)$ for $k=-N/2, \ldots, N/2-1$.

All the $\fwjp$'s together with $\widehat{\phi_{n,0}}$ form an orthonormal basis of $\CC^N$ since the supports of $\fwjp$ at different level $j$'s are disjoint. Some version of Shannon wavelets may have extra phases multiplied on these $\chi_{[-2^{n-j},-2^{n-j-1})}(k)$ and $\chi_{[2^{n-j-1},2^{n-j})}(k)$ \cite[Chapter 7.2]{mallat1999wavelet}, but we do not adopt them here. 


By orthogonality, the wavelet coefficients of signal $f$ are
\begin{equation}\label{eq: Shannon coefficients}
    \cwjp = \inner{f,\wjp} = \inner{\hat{f},\fwjp} = \left(\sum_{k=2^{n-j-1}}^{2^{n-j}-1}+\sum_{k=-2^{n-j}}^{-2^{n-j-1}-1}\right)\frac{1}{\sqrt{2^{n-j}}}e^{-2\pi i \frac{pk}{2^{n-j}}}\hat{f}(k).
\end{equation}
and \begin{equation}
    \cscale = \inner{f,\phi_{n,0}} = \inner{\hat{f},\widehat{\phi_{n,0}}} = \hat{f}(0).
\end{equation}
The Shannon wavelet transform is defined as
\begin{equation}
    \uws:\CC^N\to \CC^N: f\mapsto a = (a_{1,0},\ldots,a_{1,N/2-1},a_{2,0},\ldots,a_{2,N/4-2},\ldots,a_{n-1,0},a_{n-1,1},a_{n,0},a_{n+1,0})^T.
\end{equation}
Since the calculation \eqref{eq: Shannon coefficients} first involves a Fourier transform of $f$, we introduce
\begin{equation}
    \shf{N}:\CC^N\to \CC^N: \hat{f}\mapsto a,
\end{equation}
which satisfies $\uws = \shf{N} U_{\mathrm{FT}(N)}$. Notice that when $N=2$, $\shf{2} = X$ is the Pauli X matrix.

Consider the rearrangement of the index of $\hat{f}$ as follows:
\begin{align*}
    & \trans{N}:(\hat{f}(0),\ldots,\hat{f}(N/2-1),\hat{f}(-N/2),\ldots,\hat{f}(-1))^T\\
    \mapsto {}& (\hat{f}(-N/2),\ldots,\hat{f}(-N/4-1),\hat{f}(N/4),\ldots,\hat{f}(N/2-1),\hat{f}(0),\ldots,\hat{f}(N/4-1),\hat{f}(-N/4),\ldots,\hat{f}(-1))^T,
\end{align*}
which is the transposition of the first and the third quarter of the indices. The matrix form of $\trans{N}$ can be written explicitly as $\trans{N} = \swaponethree\otimes I_{N/4}$, where
\begin{equation*}
\begin{small}
\begin{tikzpicture}[on grid]
\pgfkeys{/myqcircuit, layer width=7.5mm, row sep=5mm, source node=qwsource}
\newcommand{\qwstart}{1.5}
\newcommand{\qwend}{1.5}
\qwire[index=1, start layer=\qwstart, end layer=\qwend, end node=qw1e]
\qwire[index=2, start layer=\qwstart, end layer=\qwend, end node=qw2e]

\renewcommand{\qwstart}{4}
\renewcommand{\qwend}{6}
\qwire[index=1, start layer=\qwstart, end layer=\qwend, start node=qw1s]
\qwire[index=2, start layer=\qwstart, end layer=\qwend, start node=qw2s]
\cnot[layer=4, control index=2, target index=1, style=controloff]
\node at ($0.5*(qw1e)+0.5*(qw2e)$) {$\swaponethree := \begin{bmatrix}
        0&0&1&0\\
        0&1&0&0\\
        1&0&0&0\\
        0&0&0&1
    \end{bmatrix}=$};
\end{tikzpicture}%
\end{small}
\end{equation*}
is just a $\ket{0}$ controlled-NOT gate.
After this rearrangement, when $j=1$, \eqref{eq: Shannon coefficients} becomes 
\begin{equation}
    \cwjpone = \sum_{k=0}^{2^{n-1}-1}\frac{1}{\sqrt{2^{n-1}}}e^{-2\pi i \frac{pk}{2^{n-1}}}(\trans{N}\hat{f})(k),
\end{equation}
which is an inverse Fourier transform of size $\frac{N}{2}$ on the first half of $\trans{N}\hat{f}$.

\begin{figure}[!ht]
	\centering
	\includegraphics[
	width =0.7\textwidth
	]{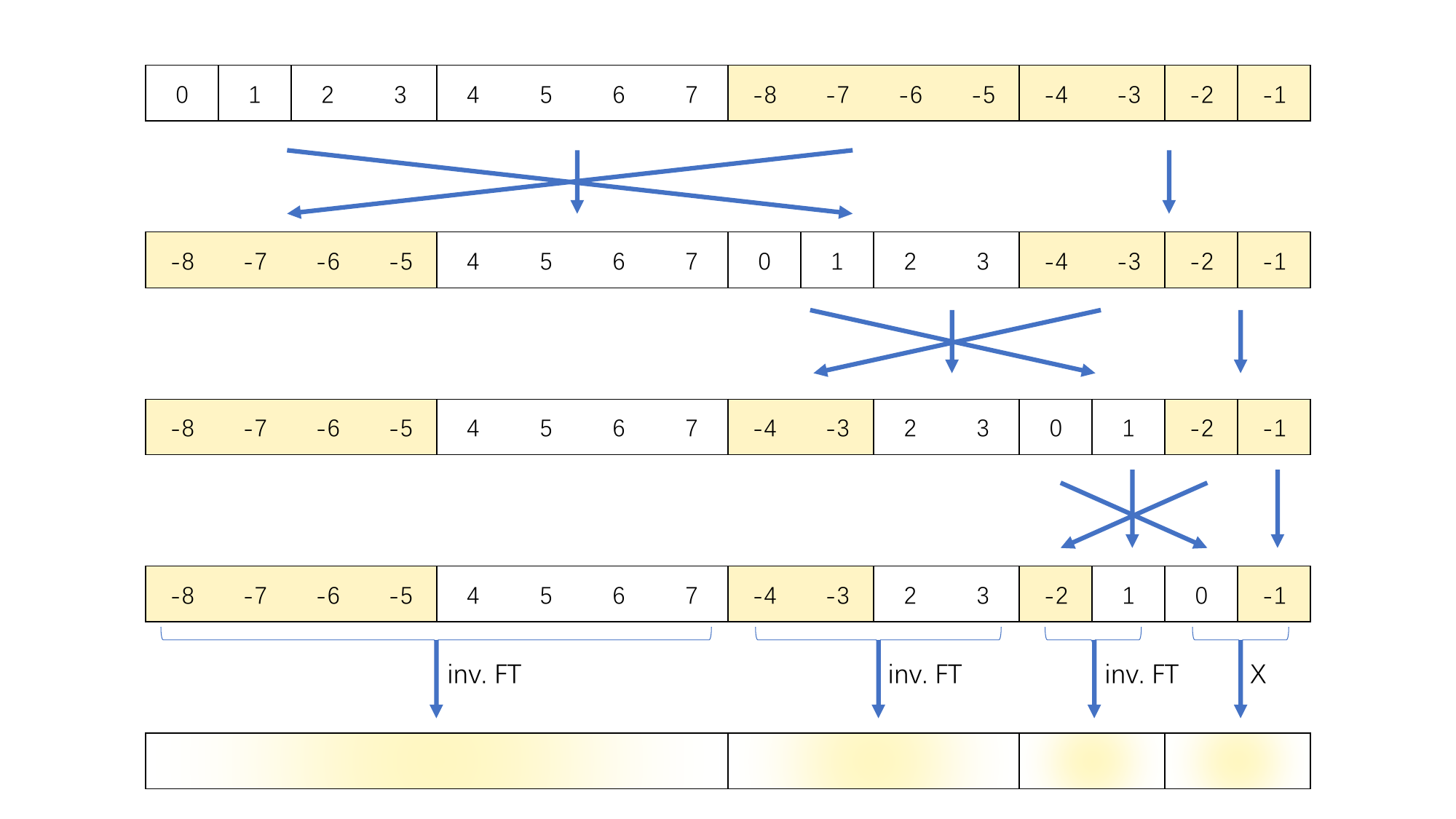}
	\caption{The illustration of $\shf{K}$. It contains a reshuffling process, followed by inverse Fourier transforms. The inverse Fourier transforms can also be performed right after the corresponding reshuffling process.}
	\label{fig: reshuffle}
\end{figure}

Moreover, the latter half of $\trans{N}\hat{f}$ is exactly the middle half of $\hat{f}$, i.e., the second and third quarters of $\hat{f}$. By definition \eqref{eq: Shannon coefficients}, we notice that the second half of the coefficient vector $a$, denoted as
\begin{equation}
    a(N/2:N-1) = (a_{2,0},\ldots,a_{2,N/4-2},\ldots,a_{n-1,0},a_{n-1,1},a_{n,0},a_{n,1})^T,
\end{equation}
only depends on $(\trans{N}\hat{f})(N/2:N-1) = \hat{f}(-N/4:N/4-1)$, and this dependence follows the same pattern as \eqref{eq: Shannon coefficients} except for $N$ is replaced by $N/2$. More specifically,
\begin{equation}
    \shf{N/2}(\trans{N}\hat{f})(N/2:N-1) = a(N/2:N-1).
\end{equation}
Therefore, $\shf{N}$ can be implemented recursively, as shown in \Cref{fig: reshuffle}, and we have
\begin{equation}
    \shf{K} = \left(\ketbra{0}{0}\otimes U_{\mathrm{FT}(K/2)}^\dagger+\ketbra{1}{1}\otimes\shf{K/2}\right)\trans{K}
\end{equation}
for $K = N, N/2, N/4,\ldots, 4$, with the base case $\shf{2} = X$ being the Pauli X matrix. This can be implemented in the quantum circuit as \Cref{fig: shfk}. The fully expanded circuit is shown in \Cref{fig: Shannon wavelet}. This circuit also contains multi-qubit controlled unitary matrices, which can be implemented as in \Cref{fig: multi-qubit control}.


\begin{figure}[htp]
\centering
\centering
\begin{small}
\begin{tikzpicture}[on grid]
\pgfkeys{/myqcircuit, layer width=7.5mm, row sep=5mm, source node=qwsource}
\newcommand{\qwstart}{1}
\newcommand{\qwend}{3}
\qwire[index=1, start layer=\qwstart, end layer=\qwend, label=$~$]
\qwire[index=2, start layer=\qwstart, end layer=\qwend, label=$~$, end node=qw2a]
\qwire[index=3, start layer=\qwstart, end layer=\qwend, label=$~$, end node=qw3a]
\qwire[index=4, start layer=\qwstart, end layer=\qwend, label=$~$, end node = qwe]
\multiqubit[layer=1, start index=1, stop index=4, label=$\,\shf{K}\,$]

\renewcommand{\qwstart}{5}
\renewcommand{\qwend}{12}
\qwire[index=1, start layer=\qwstart, end layer=\qwend]
\qwire[index=2, start layer=\qwstart, end layer=\qwend]
\qwire[index=3, start layer=\qwstart, end layer=\qwend, start node=qw3b]
\qwire[index=4, start layer=\qwstart, end layer=\qwend, end node=qwe]
\singlequbit[style=not, layer=5, index=1, node=not1]
\control[layer=5, index=2, target node=not1, style=controloff]
\multiqubit[layer=7, start index=2, stop index=4, node=iqft, label=$\,U_{\mathrm{FT}(K/2)}^\dagger\,$]
\control[layer=7, index=1, target node=iqft, style=controloff]
\multiqubit[layer=9, start index=2, stop index=4, node=sk2, label=$\,\shf{K/2}\,$]
\control[layer=9, index=1, target node=sk2, style=controlon]
\node at ($0.5*(qw3a)+0.5*(qw2a)+(1,0)$) {$=$};
\end{tikzpicture}%
\end{small}
\caption{The recursive definition of $\shf{K}$. The base case is $\shf{2} =X$.}
\label{fig: shfk}
\end{figure}
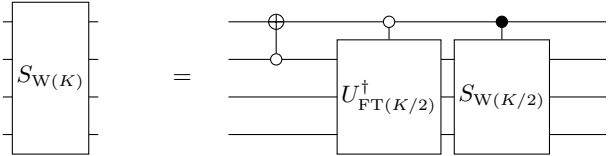

\begin{figure}[!ht]
\centering
\begin{small}
\begin{tikzpicture}[on grid]
\pgfkeys{/myqcircuit, layer width=7.5mm, row sep=5mm, source node=qwsource}
\newcommand{\qwstart}{1}
\newcommand{\qwend}{14}
\qwire[index=1, start layer=\qwstart, end layer=\qwend, label=$\ket{x_3}$]
\qwire[index=2, start layer=\qwstart, end layer=\qwend, label=$\ket{x_2}$, end node=qw2a]
\qwire[index=3, start layer=\qwstart, end layer=\qwend, label=$\ket{x_1}$, end node=qw3a]
\qwire[index=4, start layer=\qwstart, end layer=\qwend, label=$\ket{x_0}$, end node = qwe]

\multiqubit[layer=1.5, start index=1, stop index=4, label=$\,U_{\mathrm{FT}(16)}\,$]

\singlequbit[style=not, layer=3, index=1, node=not1]
\control[layer=3, index=2, target node=not1, style=controloff]

\multiqubit[layer=4.5, start index=2, stop index=4, node=iqft, label=$\,U_{\mathrm{FT}(8)}^\dagger\,$]
\control[layer=4.5, index=1, target node=iqft, style=controloff]

\singlequbit[style=not, layer=6, index=2, node=not1]
\control[layer=6, index=3, target node=not1, style=controloff]
\control[layer=6, index=1, target node=not1, style=controlon]

\multiqubit[layer=7.5, start index=3, stop index=4, node=iqft, label=$\,U_{\mathrm{FT}(4)}^\dagger\,$]
\control[layer=7.5, index=2, target node=iqft, style=controloff, node = c1]
\control[layer=7.5, index=1, target node=c1, style=controlon]

\singlequbit[style=not, layer=9, index=3, node=not1]
\control[layer=9, index=4, target node=not1, style=controloff]
\control[layer=9, index=2, target node=not1, style=controlon, node=c1]
\control[layer=9, index=1, target node=c1, style=controlon]

\multiqubit[layer=10.5, start index=4, stop index=4, node=iqft, label=$\,U_{\mathrm{FT}(2)}^\dagger\,$]
\control[layer=10.5, index=3, target node=iqft, style=controloff, node = c1]
\control[layer=10.5, index=2, target node=c1, style=controlon, node = c2]
\control[layer=10.5, index=1, target node=c2, style=controlon]

\multiqubit[layer=12, start index=4, stop index=4, node=iqft, label=$\,X\,$]
\control[layer=12, index=3, target node=iqft, style=controlon, node = c1]
\control[layer=12, index=2, target node=c1, style=controlon, node = c2]
\control[layer=12, index=1, target node=c2, style=controlon]

\end{tikzpicture}%
\end{small}%
\caption{The fully expanded circuit of $\uws$ when $N = 16$. Here, for multiple qubit control gates, the controlling qubits may be located both on the upper wires and lower wires of the controlled qubit.}
\label{fig: Shannon wavelet}
\end{figure}
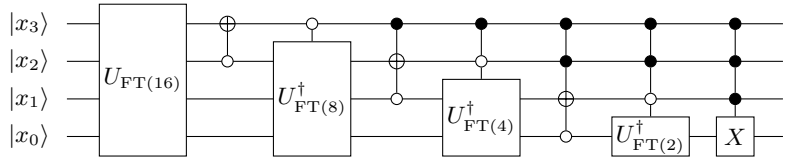

\subsection{Meyer wavelets (blended)}\label{sec: Meyer wavelet}
In the continuous setting, the Meyer wavelets are given by \eqref{eq: defi phi omega}. In order to impose a periodic boundary condition here and maintain the orthogonality, we need to introduce a phase shift and define
\begin{equation}
    \hat{\uppsi}_{ms}(\omega) = \begin{cases}
        e^{i\frac{\pi}{4}-i\frac{\omega}{2}}g\left(\frac{3\omega}{2}-2\pi\right)& \text{if $\frac{2\pi}{3}\le \omega\le\frac{4\pi}{3}$,}\\
        e^{i\frac{\pi}{4}-i\frac{\omega}{2}}g\left(\frac{3\omega}{4}-\pi\right)& \text{if $\frac{4\pi}{3}\le \omega\le\frac{8\pi}{3}$,}\\
        0& \text{if $0\le \omega\le\frac{2\pi}{3}$ or $\omega\ge\frac{8\pi}{3}$,}\\
        \hat{\uppsi}_{ms}(-\omega)^*& \text{if $\omega<0$.}\\
    \end{cases}
\end{equation}
Now we can define the discrete Meyer wavelet base functions $\wjp$ by assigning its Fourier coefficients as
\begin{equation}\label{eq: discrete Meyer basis}
    \fwjp (k) = \frac{1}{\sqrt{2^{n-j}}} e^{2\pi i\frac{pk}{2^{n-j}}}\sum_{q\in\ZZ}\hat{\uppsi}_{ms}\left(2^{j+1}\pi\left(\frac{k}{N}+q\right)\right),\quad (k = -\frac{N}{2}, -\frac{N}{2}+1,\ldots, \frac{N}{2}-1),
\end{equation}
for $j = 1,2,\ldots, n$ and $p\in[2^{n-j}]$. One can check that these $N-1$ functions, together with the scaling function
\begin{equation}
    \widehat{\phi_{n,0}} = \chi_{\{0\}}
\end{equation}
form an orthonormal basis of $\CC^{N}$. While the orthogonality can be calculated directly, it can also be seen as a corollary of the following discussions, in which we will illustrate that the expansion under this basis is unitary. When $k = -\frac{N}{2}, -\frac{N}{2}+1,\ldots, \frac{N}{2}-1$, \eqref{eq: discrete Meyer basis} can also be formulated as 
\begin{equation}
    \fwjp (k) = \begin{cases}
        \frac{1}{\sqrt{2^{n-j}}} e^{2\pi i\frac{pk}{2^{n-j}}}\sum_{q=-1}^{1}\hat{\uppsi}_{ms}\left(2^{j+1}\pi\left(\frac{k}{N}+q\right)\right) &\text{if $j=1$,}\\\frac{1}{\sqrt{2^{n-j}}} e^{2\pi i\frac{pk}{2^{n-j}}}\hat{\uppsi}_{ms}\left(2^{j+1}\pi\frac{k}{N}\right) &\text{if $j=2,3,\ldots, n$.}
    \end{cases}
\end{equation}
We also point out that if we ignore the range of $k$ in \eqref{eq: discrete Meyer basis}, it still holds $\fwjp (k) = \fwjp (k+N)$, which is compatible with our convention. Our objective is to calculate the wavelet coefficients
\begin{equation}
    \cwjp = \inner{f,\wjp} = \inner{\hat{f},\fwjp}
\end{equation}
and 
\begin{equation}
    \cscale := \inner{f,\phi_{n,0}} = \inner{\hat{f},\widehat{\phi_{n,0}}} = \hat{f}(0).
\end{equation}
Then the wavelet transform in a matrix form is \begin{equation}
    \uwb:\CC^N\to \CC^N: f\mapsto a = (a_{1,0},\ldots,a_{1,N/2-1},a_{2,0},\ldots,a_{2,N/4-2},\ldots,a_{n-1,0},a_{n-1,1},a_{n,0},a_{n+1,0})^T.
\end{equation}


\begin{figure}[!ht]
	\centering
	\includegraphics[
	width =0.7\textwidth
	]{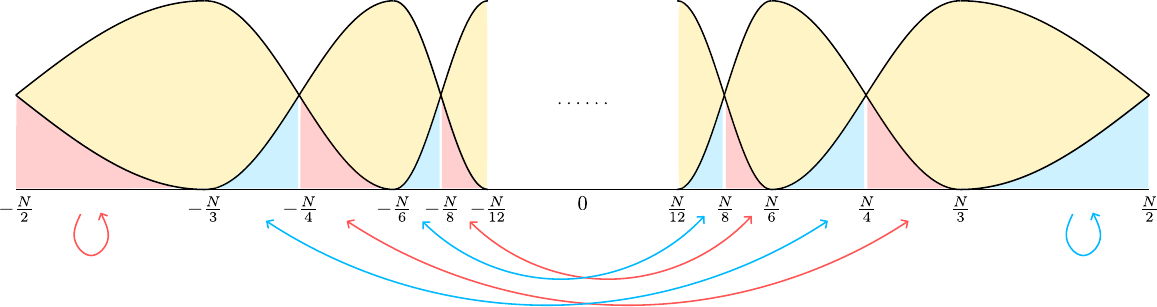}
	\caption{This figure demonstrates the reallocations of $\hat{f}(k)$ in the first few levels. $\fwjp$ is shown as two bumps supported at $[-\frac{2^{n-j+2}}{3},-\frac{2^{n-j}}{3})$ and $[\frac{2^{n-j}}{3},\frac{2^{n-j+2}}{3})$. After the transpositions, the relevant $\hat{f}(k)$'s would be placed at $[-2^{n-j},-2^{n-j-1})\cup[2^{n-j-1},2^{n-j})$ with correct proportions.}
	\label{fig: wavelet reallocate}
\end{figure}

Similar to the blended Gabor atom case, the support of basis functions at adjacent levels are overlapping. Therefore, we reallocate a portion of each $\hat{f}(k)$, which is illustrated in \Cref{fig: wavelet reallocate}. To write down this reallocation process precisely, we introduce an intermediate variable $h\in\CC^N$, which is given by
\begin{equation}\label{eq: h+n-q}
    h(\frac{N}{2^j}-q) = -e^{-i\qpi-i\frac{\pi q 2^{j}}{N}}g(\hpi-\frac{3\pi q 2^{j-1}}{N})\hat{f}(\frac{N}{2^j}-q) - e^{i\qpi-i\frac{\pi q 2^{j}}{N}}g(-\hpi-\frac{3\pi q 2^{j-1}}{N})\hat{f}(-\frac{N}{2^j}-q),
\end{equation}
\begin{equation}\label{eq: h-n-q}
    h(-\frac{N}{2^j}-q) =  e^{i\qpi-i\frac{\pi q 2^{j-1}}{N}}g(-\hpi-\frac{3\pi q 2^{j-1}}{N})\hat{f}(\frac{N}{2^j}-q) + e^{-i\qpi-i\frac{\pi q 2^{j-1}}{N}}g(\hpi-\frac{3\pi q 2^{j-1}}{N})\hat{f}(-\frac{N}{2^j}-q),
\end{equation}
for $j = 2,3,\cdots,n$ and $1\le q < \frac{N}{3\cdot 2^j}$. Similarly, we define
\begin{equation}\label{eq: h+n+q}
    h(\frac{N}{2^j}+q) = e^{i\qpi+i\frac{\pi q 2^{j-1}}{N}}g(-\hpi+\frac{3\pi q 2^{j-1}}{N})\hat{f}(\frac{N}{2^j}+q) + e^{-i\qpi+i\frac{\pi q 2^{j-1}}{N}}g(\hpi+\frac{3\pi q 2^{j-1}}{N})\hat{f}(-\frac{N}{2^j}+q),
\end{equation}
\begin{equation}\label{eq: h-n+q}
    h(-\frac{N}{2^j}+q) =  - e^{-i\qpi+i\frac{\pi q 2^{j}}{N}}g(\hpi+\frac{3\pi q 2^{j-1}}{N})\hat{f}(\frac{N}{2^j}+q) -e^{i\qpi+i\frac{\pi q 2^{j}}{N}}g(-\hpi+\frac{3\pi q 2^{j-1}}{N})\hat{f}(-\frac{N}{2^j}+q),
\end{equation}
for $j = 2,3,\cdots,n$ and $0\le q < \frac{N}{3\cdot 2^j}$.

For the first level $j=1$, there is a little bit of difference since there is no higher level and the indices $-\frac{N}{2}+q$ and $\frac{N}{2}+q$ are the same. Therefore, we only adopt \eqref{eq: h+n-q} and \eqref{eq: h-n+q} for $j=1$ and $0\le q < \frac{N}{3\cdot 2^j}$. Finally, let $h(0) = \hat{f}(0)$ for the scaling function component. We denote this transformation from $\hat{f}$ to $h$ as $\wtg$, which is $h = \wtg \hat{f}$. 

We may calculate that 
\begin{equation}\label{eq: wavelet h to a}
    \left(\sum_{k=2^{n-j-1}}^{2^{n-j}-1}+\sum_{k=-2^{n-j}}^{-2^{n-j-1}-1}\right)\frac{1}{\sqrt{2^{n-j}}}e^{-2\pi i \frac{pk}{2^{n-j}}}h(k) = \inner{\hat{f},\fwjp} = \cwjp.
\end{equation}
The proof is in \Cref{sec: proof Meyer wavelet h}. Noticing that the left-hand side of \eqref{eq: wavelet h to a} is exactly the same as applying $\shf{N}$ on $h$, we conclude
\begin{equation}\label{eq: uwm}
    \uwb = \shf{N}\wtg U_{\mathrm{FT}(N)}.
\end{equation}
Therefore, the only remaining problem is to implement the matrix $\wtg$.

\subsection{The implementation of $\wtg$}

The idea of implementing $\wtg$ is similar to $\tg$ in the Gabor atom implementation, exploiting the block diagonal structure and using the technique of encoding the exponential of pure imaginary polynomials. However, the main challenge here is the ranges of indices are not power of 2. A plausible choice to get around this difficulty is to further restrict the support of $g(x)$ to $[-\frac{7\pi}{8},\frac{7\pi}{8}]$, and also adjust the phase rotation factors $e^{-i\frac{\omega}{2}}$ in \eqref{eq: defi phi omega} in order to restrict the non-trivial part of the operations in blocks of size of power of 2. However, because this choice makes the wavelet different from what is commonly used, we choose to present a method that can implement $\wtg$ without any restrictions.

\begin{figure}[!ht]
	\centering
	\includegraphics[
	width =0.4\textwidth
	]{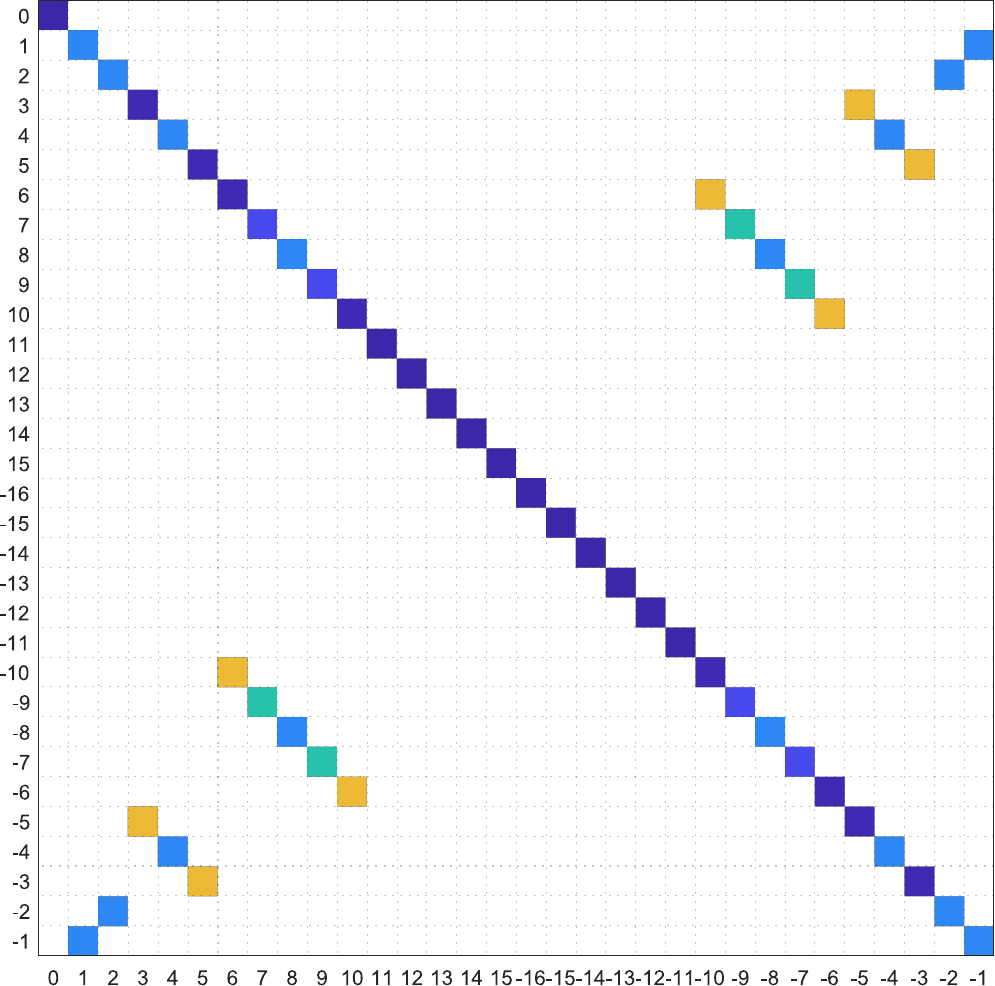}
	\caption{Magnitude of elements in $\wtg$ with $N = 32$. }
	\label{fig: wtg}
\end{figure}


From the formulas \eqref{eq: h+n-q} - \eqref{eq: h-n+q}, we see that $\wtg$ has a block-diagonal structure. This can also be seen from \Cref{fig: wtg}. Let us write the unitary block explicitly as
\begin{equation}
    \wtg = \prod_{j=1}^n \wrj{j}\prod_{j=1}^n \wlj{j}
\end{equation}
where $\wrj{j}$ is the unitary corresponding to \eqref{eq: h+n+q} and \eqref{eq: h-n+q} for level $j$, and $\wlj{j}$ is the unitary corresponding to \eqref{eq: h+n-q} and \eqref{eq: h-n-q} for level $j$. Notice that $\wrj{1}$ and $\wlj{1}$ are only defined by \eqref{eq: h-n+q} and \eqref{eq: h+n-q}, respectively. Since the support of all $\wrj{j}$'s and $\wlj{j}$'s are disjoint, they are all commutative, and thus the order of product does not matter.

\begin{figure}[!ht]
\centering
\begin{small}
\begin{tikzpicture}[on grid]
\pgfkeys{/myqcircuit, layer width=7.5mm, row sep=7mm, source node=qwsource}
\newcommand{\qwstart}{1}
\newcommand{\qwend}{6}
\qwire[index=1, start layer=\qwstart, end layer=\qwend]
\qwire[index=2, start layer=\qwstart, end layer=\qwend]
\qwire[index=3, start layer=\qwstart, end layer=\qwend]
\qwire[index=4, start layer=\qwstart, end layer=\qwend]
\qwire[index=5, start layer=\qwstart, end layer=\qwend]

\singlequbit[style=not, layer=1, index=2, node=not1]
\control[layer=1, index=1, target node=not1, style=controloff]
\singlequbit[style=not, layer=2, index=3, node=not2]
\control[layer=2, index=1, target node=not2, style=controloff]
\singlequbit[style=not, layer=3, index=4, node=not3]
\control[layer=3, index=1, target node=not3, style=controloff]
\swapgate[layer=4, first index=1, second index=5]

\end{tikzpicture}%
\end{small}%
\caption{The circuit for $\wq{32}$. }
\label{fig: wkj}
\end{figure}
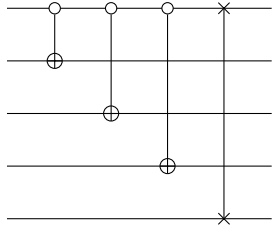

By definition, $\wrj{j}$ is acting non-trivially on the indices $\pm\frac{N}{2^j}+q$ for $0\le q<\frac{N}{3\cdot 2^j}$. However, this range is not convenient for quantum implementation since it is not a power of 2. Therefore, we may first enlarge the range of $q$ to $0\le q<\frac{N}{2^j}$ and pad the extra slots by identity. In the binary representation, this range contains the binary strings beginning with $\jzero$ and $\jone$. When $j\ge 2$, we shall first group these two ranges together and then perform the desired unitary transformation. To be specific, we need a permutation matrix $\wq{2^j}$ such that
\begin{equation}
    \wq{2^j}\ket{\jzero } = \ket{\underbrace{1\cdots1}_{(j-1)\text{ of 1's}}\!\!\!\!\!0 }\quad \text{and}\quad \wq{2^j}\ket{\jone } = \ket{\jone }.
\end{equation}
There are different choices of $\wq{2^j}$, and we adopt the following
\begin{equation}
    \begin{aligned}
    &\ket{x_{n-1}x_{n-2}\cdots x_{n-j+1}x_{n-j}}\\
    &\mapsto \ket{x_{n-1}(x_{n-2}\oplus x_{n-1}\oplus 1)\cdots (x_{n-j+1}\oplus x_{n-1}\oplus 1)(x_{n-j}\oplus x_{n-1}\oplus 1)}\\
    &\mapsto \ket{(x_{n-j}\oplus x_{n-1}\oplus 1)(x_{n-2}\oplus x_{n-1}\oplus 1)\cdots (x_{n-j+1}\oplus x_{n-1}\oplus 1)x_{n-1}},
\end{aligned}
\end{equation}
which is illustrated in \Cref{fig: wkj}.

After applying $\wq{2^j}$, the relevant indices begin with $\underbrace{1\cdots1}_{(j-1)\text{ of 1's}}\!\!\!\!\!0 $ and $\jone$. Therefore, when $j\ge 2$, the matrix $\wrj{j}$ can be written as 
\begin{equation}\label{eq: wrj greater than 2}
    \wrj{j} = (\wq{2^j}^\dagger\otimes I_{2^{n-j}})(I_{N-2^{n-j+1}} \oplus \wk{2^{n-j+1}})(\wq{2^j}\otimes I_{2^{n-j}})
\end{equation}
where $\wk{2^{n-j+1}}$ is a $2^{n-j+1}$ dimensional matrix that can be determine according to \eqref{eq: h+n+q} and \eqref{eq: h-n+q}. Let $D$ be a $\lceil\frac{N}{3\cdot 2^j}\rceil$ dimensional diagonal matrix $D = \diag\{-\half+\frac{3q2^{j-1}}{N}: 0\le q<\frac{N}{3\cdot 2^j}\}$, then
\begin{equation}
    \wk{2^{n-j+1}} = \begin{bmatrix}
        e^{i\pi(\frac{1}{3}D+\frac{5}{12})}\cos(\halfpi\beta(D))& 0&-e^{i\pi(\frac{1}{3}D-\frac{1}{12})}\sin(\halfpi\beta(D))&0\\
        0&I&0&0\\
        e^{i\pi(\frac{2}{3}D+\frac{1}{12})}\sin(\halfpi\beta(D))&0 &- e^{i\pi(\frac{2}{3}D+\frac{7}{12})}\cos(\halfpi\beta(D))&0\\
        0&0&0&I
    \end{bmatrix},
\end{equation}
where $I$ stands for the $\frac{N}{2^j}-\lceil\frac{N}{3\cdot 2^j}\rceil$ dimensional identity matrix. Note that as this matrix only depends on the value of $n-j$, it makes sense to write $\wk{2^{m+1}}$ without specifying $n$ and $j$.

When $j=1$, the matrix $\wrj{1}$ is different from the \eqref{eq: wrj greater than 2}. According to \eqref{eq: h-n+q}, we have
\begin{equation}
    \wrj{1} = I_{N/2}\oplus \widetilde{\wk{N/2}}
\end{equation}
with
\begin{equation}\label{eq: wrj1}
    \widetilde{\wk{N/2}} = \begin{bmatrix}
        e^{i\pi(-\frac{5}{12}+\frac{2}{3}D+\half\beta(D))}&0\\
        0&I
    \end{bmatrix},
\end{equation}
where $D = \diag\{-\half+\frac{3q}{N}: 0\le q<\frac{N}{6}\}$, and $I$ stands for the $\frac{N}{2}-\lceil\frac{N}{6}\rceil$ dimensional identity matrix.

\begin{figure}[!ht]
\centering
\begin{small}
\begin{tikzpicture}[on grid]
\pgfkeys{/myqcircuit, layer width=7.5mm, row sep=7mm, source node=qwsource}
\newcommand{\qwstart}{1}
\newcommand{\qwend}{16}
\qwire[index=1, start layer=2+\qwstart, end layer=\qwend -1, label = ancilla $\ket{0}$]
\qwire[index=2, start layer=1+\qwstart, end layer=\qwend-1, label = $\ket{x_4}$]
\qwire[index=3, start layer=1+\qwstart, end layer=\qwend-1, label = $\ket{x_3}$]
\qwire[index=4, start layer=1+\qwstart, end layer=\qwend-1, label = $\ket{x_2}$]
\qwire[index=5, start layer=1+\qwstart, end layer=\qwend-1, label = $\ket{x_1}$]
\qwire[index=6, start layer=1+\qwstart, end layer=\qwend-1, label = $\ket{x_0}$]

\singlequbit[style=not, layer=3, index=1, node=not1]
\control[layer=3, index=2, target node=not1, style=controloff, node=Z1]
\control[layer=3, index=3, target node=Z1, style=controloff]

\singlequbit[style=not, layer=4, index=1, node=not2]
\control[layer=4, index=2, target node=not2, style=controloff, node=Z1]
\control[layer=4, index=3, target node=Z1, style=controlon, node=Z2]
\control[layer=4, index=4, target node=Z2, style=controloff, node=Z3]
\control[layer=4, index=5, target node=Z3, style=controloff, node=Z4]

\singlequbit[style=not, layer=5, index=1, node=not3]
\control[layer=5, index=2, target node=not3, style=controloff, node=Z1]
\control[layer=5, index=3, target node=Z1, style=controlon, node=Z2]
\control[layer=5, index=4, target node=Z2, style=controloff, node=Z3]
\control[layer=5, index=5, target node=Z3, style=controlon, node=Z4]
\control[layer=5, index=6, target node=Z4, style=controloff, node=Z5]

\multiqubit[layer=8, start index=2, stop index=6, label=$\,e^{i\pi(-\frac{5}{12}+\frac{2}{3}D_p+\half\beta(D_p))}\,$, node = Y]
\control[layer=8, index=1, target node=Y, style=controlon]

\singlequbit[style=not, layer=11, index=1, node=not3]
\control[layer=11, index=2, target node=not3, style=controloff, node=Z1]
\control[layer=11, index=3, target node=Z1, style=controlon, node=Z2]
\control[layer=11, index=4, target node=Z2, style=controloff, node=Z3]
\control[layer=11, index=5, target node=Z3, style=controlon, node=Z4]
\control[layer=11, index=6, target node=Z4, style=controloff, node=Z5]

\singlequbit[style=not, layer=12, index=1, node=not2]
\control[layer=12, index=2, target node=not2, style=controloff, node=Z1]
\control[layer=12, index=3, target node=Z1, style=controlon, node=Z2]
\control[layer=12, index=4, target node=Z2, style=controloff, node=Z3]
\control[layer=12, index=5, target node=Z3, style=controloff, node=Z4]

\singlequbit[style=not, layer=13, index=1, node=not1]
\control[layer=13, index=2, target node=not1, style=controloff, node=Z1]
\control[layer=13, index=3, target node=Z1, style=controloff]

\pgfkeys{/myqcircuit, gate offset=1}
\renewcommand{\qwstart}{16}
\renewcommand{\qwend}{16}

\qwire[index=1, start layer=\qwstart, end layer=\qwend, label = $\ket{0}$]

\end{tikzpicture}%
\end{small}%
\caption{The circuit for $\widetilde{\wk{32}}$.}
\label{fig: tilde wk}
\end{figure}
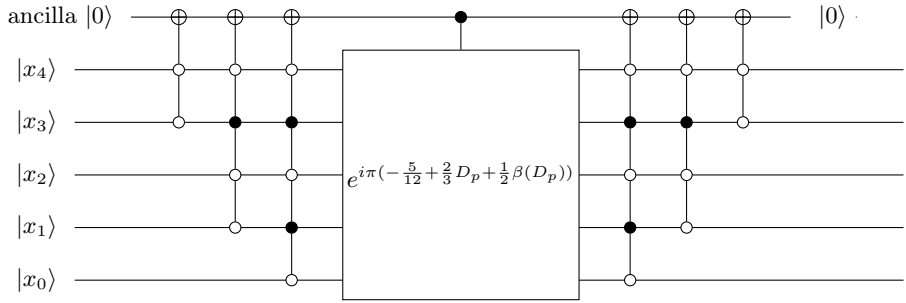

Next, we focus on implementing \eqref{eq: wrj greater than 2} and \eqref{eq: wrj1}. We first discuss the latter one since it has a more simple form. The idea is to use an ancilla qubit to filter the binary representations of numbers smaller than $\frac{N}{6}$. Recall that $\widetilde{\wk{N/2}}$ is of dimension $\frac{N}{2}$ and the largest number less than $\frac{N}{6}$ is given by the binary representation $\underbrace{0101\cdots}_{(n-1)\text{ bits}}$. Therefore, all numbers less than $\frac{N}{6}$ have the binary representation begin with $00$, or $0100$, or $010100$, or so on. Once we use control gates to encode these conditional statements into the ancilla qubit, we can perform a controlled version of $e^{i\pi(-\frac{5}{12}+\frac{2}{3}D_p+\half\beta(D_p))}$, where $D_p$ is the padded version of $D$, namely $D_p = \diag\{-\half+\frac{3q}{N}: 0\le q<\frac{N}{2}\}$. This $D_p$ can be implemented using the method discussed in \Cref{sec: exp i poly A bit}. Finally, one should go through the uncompute process to restore the ancilla to $\ket{0}$ and discard it. The complete circuit is shown in \Cref{fig: tilde wk}.

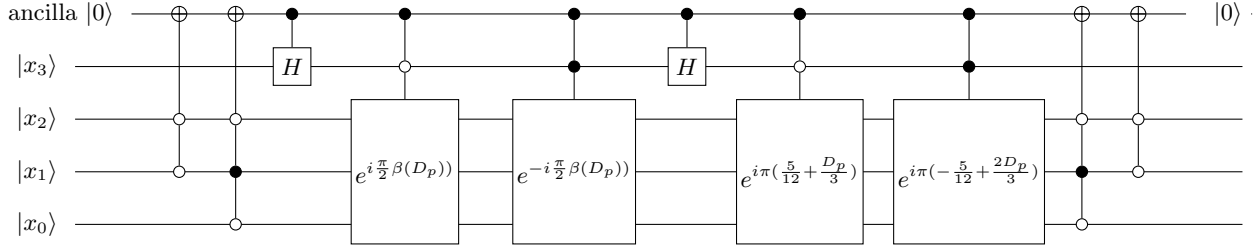
\begin{figure}[!ht]
\centering
\begin{small}
\begin{tikzpicture}[on grid]
\pgfkeys{/myqcircuit, layer width=7.5mm, row sep=7mm, source node=qwsource}
\newcommand{\qwstart}{1}
\newcommand{\qwend}{22}
\qwire[index=1, start layer=1+\qwstart, end layer=\qwend - 1, label = ancilla $\ket{0}$]
\qwire[index=2, start layer=\qwstart, end layer=\qwend, label = $\ket{x_3}$]
\qwire[index=3, start layer=\qwstart, end layer=\qwend, label = $\ket{x_2}$]
\qwire[index=4, start layer=\qwstart, end layer=\qwend, label = $\ket{x_1}$]
\qwire[index=5, start layer=\qwstart, end layer=\qwend, label = $\ket{x_0}$]

\singlequbit[style=not, layer=2, index=1, node=not1]
\control[layer=2, index=3, target node=not1, style=controloff, node=Z1]
\control[layer=2, index=4, target node=Z1, style=controloff]

\singlequbit[style=not, layer=3, index=1, node=not2]
\control[layer=3, index=3, target node=not2, style=controloff, node=Z1]
\control[layer=3, index=4, target node=Z1, style=controlon, node=Z2]
\control[layer=3, index=5, target node=Z2, style=controloff, node=Z3]

\singlequbit[style=gate, layer=4, index=2, node=H, label = $H$]
\control[layer=4, index=1, target node=H, style=controlon]

\multiqubit[layer=6, start index=3, stop index=5, label=$\,e^{i\halfpi\beta(D_p))}\,$, node = Y]
\control[layer=6, index=2, target node=Y, style=controloff, node = Z]
\control[layer=6, index=1, target node=Z, style=controlon]

\multiqubit[layer=9, start index=3, stop index=5, label=$\,e^{-i\halfpi\beta(D_p))}\,$, node = Y]
\control[layer=9, index=2, target node=Y, style=controlon, node = Z]
\control[layer=9, index=1, target node=Z, style=controlon]

\singlequbit[style=gate, layer=11, index=2, node=H, label = $H$]
\control[layer=11, index=1, target node=H, style=controlon]

\multiqubit[layer=13, start index=3, stop index=5, label=$\,e^{i\pi(\frac{5}{12}+\frac{D_p}{3})}\,$, node = Y]
\control[layer=13, index=2, target node=Y, style=controloff, node = Z]
\control[layer=13, index=1, target node=Z, style=controlon]

\multiqubit[layer=16, start index=3, stop index=5, label=$\,e^{i\pi(-\frac{5}{12}+\frac{2D_p}{3})}\,$, node = Y]
\control[layer=16, index=2, target node=Y, style=controlon, node = Z]
\control[layer=16, index=1, target node=Z, style=controlon]

\singlequbit[style=not, layer=18, index=1, node=not2]
\control[layer=18, index=3, target node=not2, style=controloff, node=Z1]
\control[layer=18, index=4, target node=Z1, style=controlon, node=Z2]
\control[layer=18, index=5, target node=Z2, style=controloff, node=Z3]

\singlequbit[style=not, layer=19, index=1, node=not1]
\control[layer=19, index=3, target node=not1, style=controloff, node=Z1]
\control[layer=19, index=4, target node=Z1, style=controloff]

\pgfkeys{/myqcircuit, gate offset=1}
\renewcommand{\qwstart}{22}
\renewcommand{\qwend}{22}

\qwire[index=1, start layer=\qwstart, end layer=\qwend, label = $\ket{0}$]

\end{tikzpicture}%
\end{small}%
\caption{The circuit for $\wk{16}$.}
\label{fig: wk}
\end{figure}

Next, let us construct the circuit of matrix $\wk{2^{n-j+1}}$. We similarly introduce the padded diagonal matrix $D_p = \diag\{-\half+\frac{3q2^{j-1}}{N}: 0\le q<\frac{N}{2^j}\}$. One may also introduce an ancilla qubit to record whether the binary number given by the last $n-j$ qubits is less than $\frac{N}{3\cdot 2^j}$. Therefore, under the control of the ancilla qubit, the matrix to be implemented is 
\begin{equation}
    \begin{bmatrix}
        e^{i\pi(\frac{1}{3}D_p+\frac{5}{12})}\cos(\halfpi\beta(D_p))&-e^{i\pi(\frac{1}{3}D_p-\frac{1}{12})}\sin(\halfpi\beta(D_p))\\
        e^{i\pi(\frac{2}{3}D_p+\frac{1}{12})}\sin(\halfpi\beta(D_p)) &- e^{i\pi(\frac{2}{3}D_p+\frac{7}{12})}\cos(\halfpi\beta(D_p))
    \end{bmatrix},
\end{equation}
which can be further decomposed as
\begin{equation}
    \begin{bmatrix}
        e^{i\pi(\frac{1}{3}D_p+\frac{5}{12})}&0\\
        0&e^{i\pi(\frac{2}{3}D_p-\frac{5}{12})}
    \end{bmatrix}
    \begin{bmatrix}
        \frac{I}{\sqrt{2}}&\frac{I}{\sqrt{2}}\\
        \frac{I}{\sqrt{2}}&-\frac{I}{\sqrt{2}}
    \end{bmatrix}
    \begin{bmatrix}
        e^{\frac{i\pi}{2}\beta(D_p)}&0\\
        0&e^{-\frac{i\pi}{2}\beta(D_p)}
    \end{bmatrix}
    \begin{bmatrix}
        \frac{I}{\sqrt{2}}&\frac{I}{\sqrt{2}}\\
        \frac{I}{\sqrt{2}}&-\frac{I}{\sqrt{2}}
    \end{bmatrix}.
\end{equation}
These four unitary matrices are either Hadamard gates or the exponentials of pure imaginary polynomials, which are constructed in \Cref{sec: exp i poly A bit}. Therefore, we only need to perform the controlled version of them and finally do the uncomputation. The complete circuit is shown in \Cref{fig: wk}.

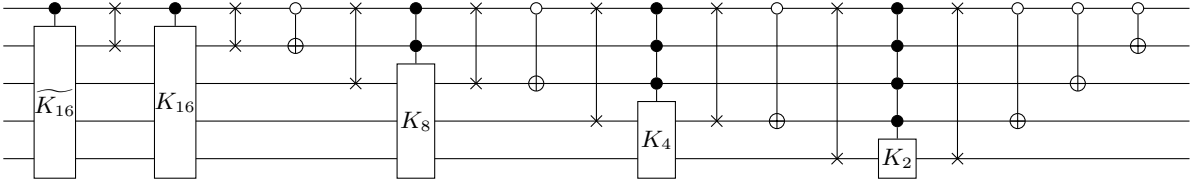
\begin{figure}[!ht]
\centering
\begin{small}
\begin{tikzpicture}[on grid]
\pgfkeys{/myqcircuit, layer width=8mm, row sep=5mm, source node=qwsource}
\newcommand{\qwstart}{1}
\newcommand{\qwend}{21}
\qwire[start node=qw1s, end node=qw1e, style=thin, index=1, start layer=\qwstart, end layer=\qwend, label=$~$]
\qwire[start node=qw2s, end node=qw2e, style=thin, index=2, start layer=\qwstart, end layer=\qwend, label=$~$]
\qwire[start node=qw3s, end node=qw3e, style=thin, index=3, start layer=\qwstart, end layer=\qwend, label=$~$]
\qwire[start node=qw1s, end node=qw1e, style=thin, index=4, start layer=\qwstart, end layer=\qwend, label=$~$]
\qwire[start node=qw2s, end node=qw2e, style=thin, index=5, start layer=\qwstart, end layer=\qwend, label=$~$]

\multiqubit[layer=1, start index=2, stop index=5, label=$\widetilde{\wk{16}}$, node = twk8]
\control[layer=1, index=1, target node=twk8, style=controlon]

\swapgate[layer=2, first index=1, second index=2]

\multiqubit[layer=3, start index=2, stop index=5, label=${\wk{16}}$, node = wk8]
\control[layer=3, index=1, target node=wk8, style=controlon]

\swapgate[layer=4, first index=1, second index=2]
\singlequbit[style=not, layer=5, index=2, node=not1]
\control[layer=5, index=1, target node=not1, style=controloff, node=Z1]
\swapgate[layer=6, first index=1, second index=3]

\multiqubit[layer=7, start index=3, stop index=5, label=${\wk{8}}$, node = wk4]
\control[layer=7, index=2, target node=wk4, style=controlon, node = n1]
\control[layer=7, index=1, target node=n1, style=controlon, node = n2]

\swapgate[layer=8, first index=1, second index=3]
\singlequbit[style=not, layer=9, index=3, node=not1]
\control[layer=9, index=1, target node=not1, style=controloff, node=Z1]
\swapgate[layer=10, first index=1, second index=4]

\multiqubit[layer=11, start index=4, stop index=5, label=${\wk{4}}$, node = wk2]
\control[layer=11, index=3, target node=wk2, style=controlon, node = n1]
\control[layer=11, index=2, target node=n1, style=controlon, node = n2]
\control[layer=11, index=1, target node=n2, style=controlon, node = n3]

\swapgate[layer=12, first index=1, second index=4]
\singlequbit[style=not, layer=13, index=4, node=not1]
\control[layer=13, index=1, target node=not1, style=controloff, node=Z1]
\swapgate[layer=14, first index=1, second index=5]

\multiqubit[layer=15, start index=5, stop index=5, label=${\wk{2}}$, node = wk2]
\control[layer=15, index=4, target node=wk2, style=controlon, node = n1]
\control[layer=15, index=3, target node=n1, style=controlon, node = n2]
\control[layer=15, index=2, target node=n2, style=controlon, node = n3]
\control[layer=15, index=1, target node=n3, style=controlon, node = n4]

\swapgate[layer=16, first index=1, second index=5]
\singlequbit[style=not, layer=17, index=4, node=not1]
\control[layer=17, index=1, target node=not1, style=controloff, node=Z1]
\singlequbit[style=not, layer=18, index=3, node=not1]
\control[layer=18, index=1, target node=not1, style=controloff, node=Z1]
\singlequbit[style=not, layer=19, index=2, node=not1]
\control[layer=19, index=1, target node=not1, style=controloff, node=Z1]

\end{tikzpicture}%
\end{small}%
\caption{The circuit of $\prod_{j=1}^n \wrj{j}$ when $N =32$.}
\label{fig: prod of R_j}
\end{figure}

Having described the implementation of all the building blocks, we conclude by drawing the circuit of $\prod_{j=1}^n \wrj{j} = \wrj{n}\wrj{n-1}\cdots\wrj{1}$ in \Cref{fig: prod of R_j}. Note that there would be cancellations at the consecutive $(\wq{2^{j+1}}^\dagger\otimes I_{2^{n-j-1}})(\wq{2^j}\otimes I_{2^{n-j}})$, which saves some gates.

For the implementation of $\prod_{j=1}^n \wlj{j}$, the method of implementation is very similar to $\prod_{j=1}^n \wrj{j}$. Therefore, we omit the details.

\subsection{Complexity analysis}
Here, we briefly discuss the complexity of the quantum wavelet transform circuits. For the exact implementation, the complexity is of $O(\mathrm{poly}(n))$, and the degree may depend on the degree of $\beta$, which is similar to the Gabor atom case. 

If we allow for an $\eps$ error, it is possible to reduce the complexity as in the Gabor atom case. In particular, we may use the inexact Fourier transform \cite{nam2020approximate}, implement the multi-qubit control by \Cref{fig: multi-qubit control}, and implement the required diagonal matrices using the method in \Cref{sec: exp i poly A qsvt}. For the Shannon wavelet, the corresponding complexity is $O(n^2\log(n\eps^{-1}))$ since we need to perform $n$ (inverse) Fourier transforms. For the Meyer wavelet, we claim that the complexity is $O(n^2(\log n+\log(\eps^{-1})^2))$ by examining each term in \eqref{eq: uwm} as follows. The $\shf{N}$ is of complexity $O(n^2\log(n\eps^{-1}))$ as discussed in the Shannon wavelet. The $\wtg$ part is mainly composed of implementing $O(n)$ diagonal matrices, and each of them costs $O(n\log(\eps^{-1})^2)$ as discussed in \Cref{sec: exp i poly A qsvt}.

Notice that for the implementation of a required diagonal matrix, one needs an ancilla qubit, as shown in \Cref{fig: tilde wk} and \Cref{fig: wk}. At the same time, one extra ancilla qubit is needed for unwrapping the multi-qubit control on the diagonal matrices. Therefore, in total two extra ancilla qubits are required.

\section{Proofs}

\subsection{Proof of \Cref{eq: Meyer packet h}}\label{sec: proof Meyer packet h}
We may further split the sum of the left-hand side into four terms
\begin{equation}\label{eq: wave packet cal left}
    \left(
    \sum_{k\in[jB,(j+\half)B)}
    +\sum_{k\in[(j+\half)B,(j+1)B)}
    +\sum_{k\in[-(j+1)B,-(j+\half)B)}
    +\sum_{k\in[-(j+\half)B,-jB)}
    \right)\frac{1}{\sqrt{2B}}e^{-2\pi i \frac{pk}{2B}}h(k).
\end{equation}
The right-hand side is the sum on the support of $\fpjp$
\begin{equation}\label{eq: wave packet cal right}
    \begin{aligned}
    & \inner{\hat{f},\fpjp} = \left(\sum_{k\in[(j-\half)B,(j+\frac{3}{2})B)} + \sum_{k\in[-(j+\frac{3}{2})B,-(j+\half)B)}\right)\hat{f}(k)\fpjp(k)^*\\
    =& \left(
    \sum_{\substack{k\in[jB,(j+\half)B)\\ \cup [-jB,(-j+\half)B)}}
    +\sum_{\substack{k\in[(j+\half)B,(j+1)B)\\ \cup [-(j+\frac{3}{2})B,-(j+1)B)}}
    +\sum_{\substack{k\in[-(j+1)B,-(j+\half)B)\\ \cup [(j+1)B,(j+\frac{3}{2})B)}}
    +\sum_{\substack{k\in[-(j+\half)B,-jB)\\ \cup [(j-\half)B,jB)}}
    \right)\hat{f}(k)\fpjp(k)^*.
    \end{aligned}
\end{equation}
Using the definition of $h$, we can check that the four summations in \eqref{eq: wave packet cal left} and \eqref{eq: wave packet cal right} are equal, respectively. For instance, the first summation
\begin{equation}
    \begin{aligned}
    &{}\sum_{\substack{k\in[jB,(j+\half)B)\\ \cup [-jB,(-j+\half)B)}}\hat{f}(k)\fpjp(k)^*\\
    =&{}\sum_{q\in[0,\frac{B}{2})}\hat{f}(jB+q)\fpjp(jB+q)^*
    + \sum_{q\in[0,\frac{B}{2})}\hat{f}(-jB+q)\fpjp(-jB+q)^*\\
    =&{}\sum_{q\in[0,\frac{B}{2})}\hat{f}(jB+q)\frac{1}{\sqrt{2B}}e^{-2\pi i \frac{p(jB+q)}{2B}}e^{\halfi(-\hlfpi+q\frac{\pi}{B})}g\left(-\halfpi+q\frac{\pi}{B}\right)\\
    &{}+ \sum_{q\in[0,\frac{B}{2})}\hat{f}(-jB+q)\frac{1}{\sqrt{2B}}e^{-2\pi i \frac{p(-jB+q)}{2B}}e^{\halfi(\hlfpi+q\frac{\pi}{B})}g\left(\halfpi+q\frac{\pi}{B}\right)\\  
    =&{}\sum_{q\in[0,\frac{B}{2})}\frac{1}{\sqrt{2B}}e^{-2\pi i \frac{p(jB+q)}{2B}}h(jB+q)\qquad \text{(used \eqref{eq: defvg1})}\\
    =&{}\sum_{k\in[jB,(j+\half)B)}
    \frac{1}{\sqrt{2B}}e^{-2\pi i \frac{pk}{2B}}h(k).
    \end{aligned}
\end{equation}
The other three summations can be checked similarly. Therefore, we have proved \Cref{eq: Meyer packet h}.

\subsection{Proof of \Cref{eq: wavelet h to a}}\label{sec: proof Meyer wavelet h}
To verify this, we may further split the sum of the left-hand side into
\begin{equation}\label{eq: wavelet cal left}
    \left(
    \sum_{k\in[-\frac{4}{3} 2^{n-j-1}, -2^{n-j-1})}
    +\sum_{k\in[-2^{n-j},-\frac{2}{3} 2^{n-j})}
    +\sum_{k\in[\frac{2}{3} 2^{n-j},2^{n-j})}
    +\sum_{k\in[2^{n-j-1},\frac{4}{3} 2^{n-j-1})}
    \right)\frac{1}{\sqrt{2^{n-j}}}e^{-2\pi i \frac{pk}{2^{n-j}}}h(k).
\end{equation}
The right-hand side is the sum on the support of $\fwjp$
\begin{equation}\label{eq: wavelet cal right}
    \begin{aligned}
    & \inner{\hat{f},\fwjp} = \left(\sum_{k\in[\frac{2}{3}2^{n-j-1},\frac{4}{3}2^{n-j})} + \sum_{k\in[-\frac{4}{3}2^{n-j},-\frac{2}{3}2^{n-j-1})}\right)\hat{f}(k)\fwjp(k)^*\\
    =& \left(
    \sum_{\substack{k\in[-\frac{4}{3} 2^{n-j-1}, -2^{n-j-1})\\ \cup [\frac{2}{3} 2^{n-j-1}, 2^{n-j-1})}}
    +\sum_{\substack{k\in[-2^{n-j},-\frac{2}{3} 2^{n-j})\\ \cup [2^{n-j},\frac{4}{3} 2^{n-j})}}
    +\sum_{\substack{k\in[\frac{2}{3} 2^{n-j},2^{n-j})\\ \cup [-\frac{4}{3}2^{n-j},-2^{n-j})}}
    +\sum_{\substack{k\in[2^{n-j-1},\frac{4}{3} 2^{n-j-1})\\ \cup [-2^{n-j-1},-\frac{2}{3}2^{n-j-1})}}
    \right)\hat{f}(k)\fwjp(k)^*.
    \end{aligned}
\end{equation}
We shall show the four summations in \eqref{eq: wavelet cal left} and \eqref{eq: wavelet cal right} are equal, respectively. For the first summation, we have
\begin{equation}
    \begin{aligned}
    &{}\sum_{\substack{k\in[-\frac{4}{3} 2^{n-j-1}, -2^{n-j-1})\\ \cup [\frac{2}{3} 2^{n-j-1}, 2^{n-j-1})}}\hat{f}(k)\fwjp(k)^*\\
    =&{}\sum_{q\in(0,\frac{1}{3}2^{n-j-1}]}\hat{f}(-2^{n-j-1}-q)\fwjp(-2^{n-j-1}-q)^*
    + \sum_{q\in(0,\frac{1}{3}2^{n-j-1}]}\hat{f}(2^{n-j-1}-q)\fwjp(2^{n-j-1}-q)^*\\
    =&{}\sum_{q\in(0,\frac{1}{3}2^{n-j-1}]}\frac{1}{\sqrt{2^{n-j}}}\hat{f}(-2^{n-j-1}-q)e^{2\pi i p(-2^{n-j-1}-q)/2^{n-j}}\hat{\psi}_{ms}(2^{j+1-n}\pi(-2^{n-j-1}-q))^*\\
    &{}+ \sum_{q\in(0,\frac{1}{3}2^{n-j-1}]}\frac{1}{\sqrt{2^{n-j}}}\hat{f}(2^{n-j-1}-q)e^{2\pi i p(2^{n-j-1}-q)/2^{n-j}}\hat{\psi}_{ms}(2^{j+1-n}\pi(2^{n-j-1}-q))^*\\
    =&{}\sum_{q\in(0,\frac{1}{3}2^{n-j-1}]}\frac{1}{\sqrt{2^{n-j}}}\hat{f}(-2^{n-j-1}-q)e^{2\pi i p(-2^{n-j-1}-q)/2^{n-j}}e^{-i\frac{\pi}{4}-i\frac{\pi q 2^j}{N}}g(\hpi-\frac{3\pi q 2^{j}}{N})\\
    &{}+ \sum_{q\in(0,\frac{1}{3}2^{n-j-1}]}\frac{1}{\sqrt{2^{n-j}}}\hat{f}(2^{n-j-1}-q)e^{2\pi i p(2^{n-j-1}-q)/2^{n-j}}e^{i\frac{\pi}{4}-i\frac{\pi q 2^j}{N}}g(-\hpi-\frac{3\pi q 2^{j}}{N})\\
    =&{}\sum_{q\in(0,\frac{1}{3}2^{n-j-1}]}\frac{1}{\sqrt{2^{n-j}}}e^{2\pi i p(-2^{n-j-1}-q)/2^{n-j}}h(-2^{n-j-1}-q)\\
    =&{}\sum_{k\in[-\frac{4}{3} 2^{n-j-1}, -2^{n-j-1})}
    \frac{1}{\sqrt{2^{n-j}}}e^{-2\pi i \frac{pk}{2^{n-j}}}h(k).
    \end{aligned}
\end{equation}
Therefore, the first summation in \eqref{eq: wavelet cal left} and \eqref{eq: wavelet cal right} are equal. The remaining three summations are also equal, respectively, by a similar argument. Thus, we have proved \eqref{eq: wavelet h to a}.

\section{Conclusion and Discussions}

This paper presents the quantum circuit implementation of wave packets, such as Gabor atoms and wavelets, with compact frequency support. The implementations for those with sharp frequency windows are rather straightforward, with the help of quantum Fourier transform. For those with blended frequency windows, our method first reallocates the Fourier coefficients to the correct location in the frequency domain, followed by applying the transformation for the sharp-windowed versions.

There are many other kinds of wave packets, such as wave atoms \cite{demanet2007wave} and curvelets \cite{candes2005curvelet}, which give tiling in space-frequency diagram different from \Cref{fig: tile}. Our method can be extended to accommodate these wave packets as well. Another possible future direction is about higher-dimensional wave packet transforms, which could find applications in solving wave equations or the Schrödinger equation. While 2D or 3D versions of compactly supported wavelets have been discussed \cite{Li2022multilevel, Li2023threedimensional}, little has been explored regarding Meyer-type wavelets or wave packets in higher dimensions.

Another more detailed possible improvement is in the implementation of the unitary diagonal matrices discussed in \Cref{sec: exp i poly A qsvt}. We anticipate the complexity can be reduced to $O(n\log(\eps^{-1}))$. Additionally, considering that QSVT can potentially handle complex polynomials directly rather than splitting them into real and imaginary parts and using LCU to combine them \cite{GilyenSuLowEtAl2018}, it may be feasible to find a polynomial that can approximate the even function $e^{-\frac{i\pi}{4}\gamma(\arcsin(x))}$ directly while satisfying the conditions in \cite[Theorem 4]{GilyenSuLowEtAl2018}. The QET-U technique \cite{dong2022ground} may also help since we start from a unitary matrix $e^{iD_{-}}$.

\bibliographystyle{amsalpha}
\bibliography{ref}	

\end{document}